\documentclass[review]{elsarticle}

\usepackage{lineno,hyperref}
\modulolinenumbers[5]

\usepackage{amssymb}
\usepackage[cm]{fullpage}
\usepackage{caption, subcaption}
\usepackage{graphicx}
\usepackage{setspace}

\journal{Icarus}

\begin{document}

\begin{frontmatter}



\title{A New Analysis of {\it Spitzer} Observations of Comet 29P/Schwassmann-Wachmann 1}


\author[1]{Charles A Schambeau}
\author[1]{Yanga R. Fern\'andez}
\author[4]{Carey M. Lisse}
\author[2]{Nalin Samarasinha}
\author[3]{Laura M. Woodney}

\address[1]{Department of Physics, University of Central Florida, Orlando, FL 32816, USA}
\address[4]{Johns Hopkins University Applied Physics Laboratory, Laurel, MD 20723}
\address[2]{Planetary Science Institute, Tucson, AZ 85719, USA}
\address[3]{Department of Physics, California State University San Bernardino, San Bernardino, CA 92407}

\begin{abstract}

\singlespacing

\noindent We present a new analysis of {\it Spitzer} observations of comet 29P/Schwassmann-Wachmann 1 taken on UT 2003 November 21, 23, and 24, similar to a previous investigation of the observations (Stansberry et al. 2004), but using the most recent {\it Spitzer} data pipeline products and intensive image processing techniques. Analysis of images from the IRAC 5.8 \& 8.0 $\mu$m bands and the MIPS 24.0 \& 70.0 $\mu$m bands resulted in photometry measurements of the nucleus after a suite of coma modeling and removal processes were implemented.  SW1 was not identified in the 5.8 $\mu$m image from the previous work so its incorporation into this analysis is entirely new. Using the Near Earth Asteroid Thermal Model (Harris 1998) resulted in a nucleus radius measurement of $R$ = 30.2 $^{+3.7}_{-2.9}$ km and an infrared beaming parameter value of $\eta = 0.99$ $^{+0.26}_{-0.19}$. We also measured an infrared geometric albedo, $p_{5.8}$ = 0.5 $\pm$ 0.5. Extrapolating a 0.04 V-band albedo and using a normalized reflectivity gradient $S' = 14.94 \pm 1.09$ [\% (1000 \r{A})$^{-1}$] (Duffard et al. 2014) we recover an infrared albedo of $p_{5.8}$ = 0.31 in the near infrared consistent with the value recovered from thermal modeling. The dust composition extracted from IRS spectra are very comet-like, containing mainly amorphous ferromagnesian silicates (but with a minority of crystalline silicates as well), water ice, and metal sulfides.

\end{abstract}


\end{frontmatter}


\section{Introduction}

\singlespacing

Comet 29P/Schwassmann-Wachmann 1 (SW1) is a unique comet and excellent observational target to investigate for improvements in our understanding of cometary nuclei and their activity drivers. SW1 is in a nearly circular orbit with eccentricity {\it e} = 0.041830, semi-major axis {\it a} = 6.009674 AU, and inclination {\it i} = 9.3763$^{\circ}$ (orbital elements from IAU Minor Planet Center, Epoch 2014 Dec. 9.0 TT = JDT 2457000.5). The orbital properties  of SW1 also classify it as a Centaur (Jewitt 2009). With a nearly constant background level coma (believed to be driven by the supervolatile CO, with observations of CO productions rates made by Festou et al. (2001), Gunnarsson et al. (2001), and Paganini et al. (2013)), SW1 undergoes frequent outbursts in activity with an increase in magnitude ranging over several orders of magnitude (Gronkowski 2014; Kossacki and Szutowicz 2013; Trigo-Rodriguez et al. 2010). The combination of SW1's orbit and nearly constant insolation raises the question of what activity drivers are causing the outbursts. With a seemingly stable thermal environment why would the nucleus experience frequent excursions from a steady state of activity? This puzzling behavior is why SW1 is an excellent target for continued study and analysis.

The radius distribution of the Centaur population is still an area of active research. Due to their large heliocentric distances, observational size measurements are challenging. A radius range of 2 km to 41 km has been observed for Centaurs by Jewitt (2009) and the first discovered Centaur 2060 Chiron has a 210$^{+11}_{-10}$ km radius (Fornasier et al. 2013). Where does SW1 lie in this distribution? Several groups past and present have studied SW1 in order to measure properties of the comet's nucleus. Due to the continuous activity of SW1 producing a persistent coma, direct measurements of the nucleus are currently unattainable. Further, observational measurements of SW1 are hindered by its large heliocentric distance. Radius measurements of SW1 show a large range: 20.0 $\pm$ 3.0 km (Cruikshank \& Brown 1983), 8.6 $\pm$ 0.1 km (Meech et al. 1993), 27.0 $\pm$ 5.0 km (Stansberry et al. 2004), 18.7$^{+5.7}_{-5.9}$ km (Stansberry et al. 2008). This large spread of radius values shows our lack of knowledge of basic physical properties of SW1's nucleus. Using {\it Spitzer} thermal images we have made an attempt at constraining the nucleus' effective radius, infrared beaming parameter, and albedo. Using {\it Spitzer} spectral observations, we have also constrained the coma's dust composition.

The organization of the paper is as follows. Section 2 gives an overview of the {\it Spitzer} observations, including details pertaining to improvements of the {\it Spitzer} BCD pipeline and emphasizing the improvements of the presented results over their 2004 counterparts. The coma modeling and removal procedure is described in Section 3 along with an application of a Near Earth Asteroid Thermal Model (NEATM, Harris 1998) to the extracted nuclear infrared photometry measurements. In Section 4, results from IRS spectrum analysis are presented. Section 5 gives an overview of the results and their implications on the thermal evolution experienced during the dynamical evolution of SW1.    

\section{Observations}

{\it Spitzer} observations of SW1 were acquired on UT 2003 November 21, 23, and 24 during the In-Orbit Checkout (IOC)/Science Verification (SV) period (Werner et al. 2004). Table 1 gives the details of each observation (similar to Table 2 from Stansberry et al. 2004). All three instruments on {\it Spitzer} were used for the observations: Infrared Array Camera (IRAC) (Fazio et al. 2004), Infrared Spectrograph (IRS) (Houck et al. 2004), and Multiband Imaging Photometer for {\it Spitzer} (MIPS) (Rieke et al. 2004). The details of each observation are given in the following three subsections.

\begin{center}
\begin{table}
\caption{}
\resizebox{1.0\textwidth}{!}{
	\begin{tabular}{ l  c c  c  c  c }
		\multicolumn{6}{ c }{SW1 Observation Summary}\\
		\hline \hline
		Instrument	&	AORKEY		&	Epoch$^{\textrm{a}}$	&	Band$^{\textrm{b}}$	&	${t_{exposure}}^{\textrm{c}}$	&	$\Delta^{\textrm{d}}$ \\ \hline
		IRAC		&	6068736		&	21, 06:03				&	5.8				&	150						&	5.51				\\
					&				&						&	8.0				&	150						&					\\		
		IRS			&	6068992		&	23, 07:18				&	SL1				&	122						&	5.54				\\
					&				&						&	LL1, LL2			&	59						&					\\
		MIPS		&	7864064		&	24, 15:05				&	24.0				&	66						&	5.56				\\
					&				&						&	70.0				&	40						&					\\
 \hline
	\end{tabular}

}
\caption*{
\resizebox{1.0\textwidth}{!}{
	\begin{tabular}{l}
		$^{\textrm{a}}$ Day of 2003 November, UT at beginning of observations. \\
		$^{\textrm{b}}$ Imaging wavelength ($\mu$m) or spectral band.\\
		$^{\textrm{c}}$ Total exposure time per pixel for each image frame, seconds.\\
		$^{\textrm{d}}$ SW1--{\it Spitzer} distance (AU). Heliocentric distance was 5.73 AU, solar phase angle was 10.0$^{\circ}$ and the tracking rate was 8$''$.6 hr$^{-1}.$\\
	\end{tabular}
}	

}

\end{table}
\end{center}

\subsection{IRAC}

The Infrared Array Camera (IRAC) is a multi-channel infrared camera capable of imaging in 3.6, 4.5, 5.8 and 8.0 $\mu$m bands. A detailed description of the IRAC can be found in the IRAC Handbook (IRAC Instrument and Instrument Support Teams 2013). Each band images a field of view (FOV) of 5.2$'$$\times$ 5.2$'$ and a pixel scale of 1.2$''$/pixel. 

The IRAC was used in high dynamic range (HDR) mode during the observations. Unfortunately, a star was close to SW1 in the FOV producing a diffraction spike in the images. This rendered the 3.6 $\mu$m and 4.5 $\mu$m observations useless other than to acquire upper limit photometry. Of importance is the identification of SW1 in the 5.8 $\mu$m band, which had not been used in the previous analysis (Stansberry et al. 2004). The 5.8 and 8.0 $\mu$m observations entailed acquiring five 30 second exposures of SW1. The software MOPEX  (MOsaicking and Point-source EXtraction, Makovoz et al. 2012) was used to calibrate and stack the basic calibrated data images (BCDs) acquired from the {\it Spitzer} Heritage Archive, generated from version S18.25.0 of the {\it Spitzer} pipeline. The final stacked, calibrated, and cropped 5.8 $\mu$m and 8.0 $\mu$m images are shown in Figure 1. Each image has an effective exposure time of 150 seconds.

\begin{figure}
        \centering
        \begin{subfigure}[b]{0.25\textwidth}
                \includegraphics[scale = 2.0]{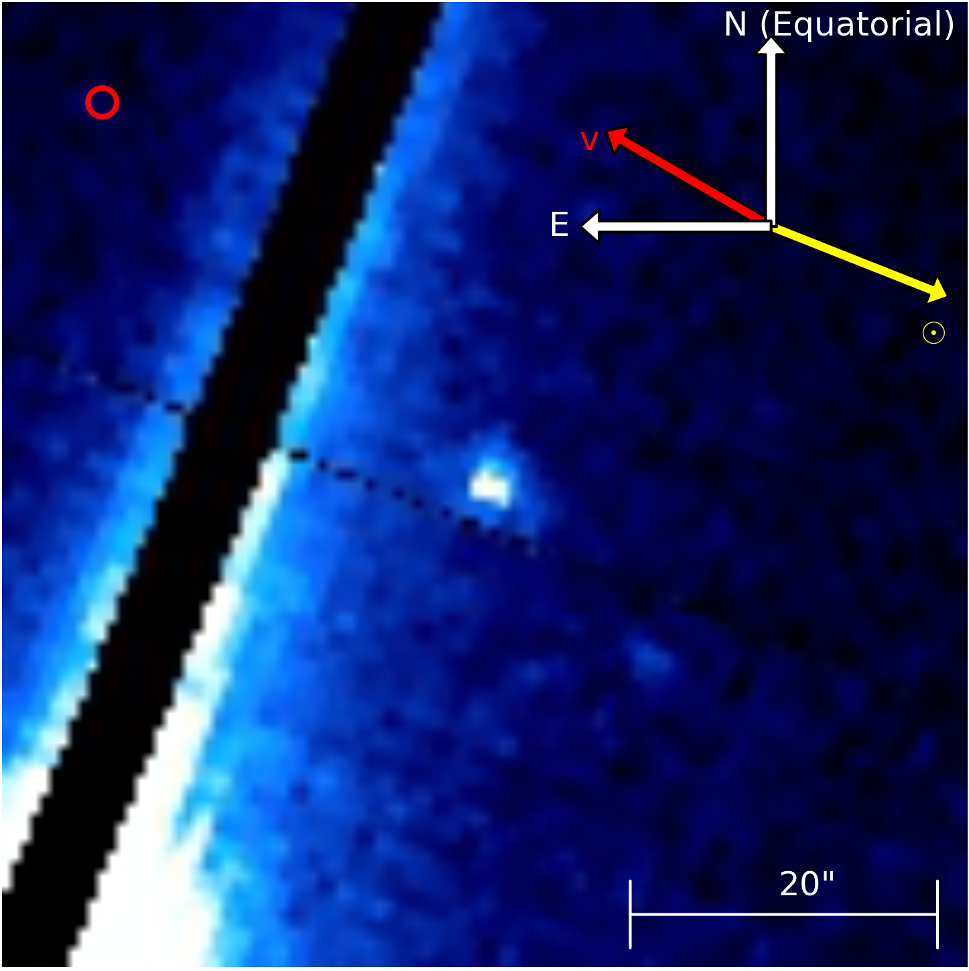}
                \caption{}
                \label{fig:comet}
        \end{subfigure}%
        ~~~~~~~~~~~~~~~~~~~~~~~~~~
        \begin{subfigure}[b]{0.25\textwidth}
                \includegraphics[scale = 2.0]{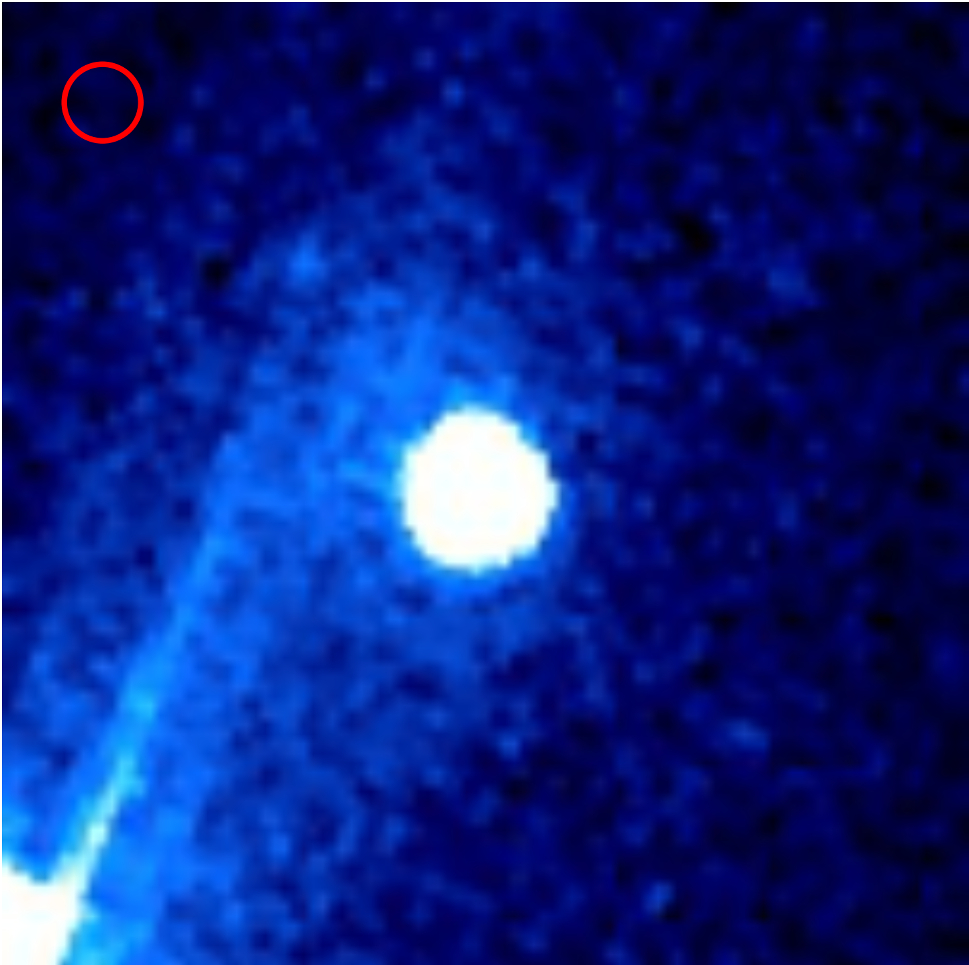}
                \caption{}
                \label{fig:tiger}
        \end{subfigure}
        \caption{(a) 5.8 $\mu$m and (b) 8.0 $\mu$m combined and calibrated images centered on SW1. Each has a FOV of  1.01$'$ $\times$ 1.01$'$ and equatorial north is up and east is to the left. The solar direction is identified by the yellow arrow and SW1's projected heliocentric velocity by the red arrow. The scale bar at the bottom right of (a) shows a distance of 20$''$ and the circle in the top left of the images represents the size of the PSF's first minimum. A star's diffraction spike artifact is clearly seen in the 5.8 $\mu$m image and partially visible in the 8.0 $\mu$m image.}
\end{figure}

In the 5.8 $\mu$m image, SW1 was identified close to the mentioned diffraction spike. A broad diffuse coma was not observed in this band. On the other hand the 8.0 $\mu$m image shows traces of a diffuse coma in the northeast and southeast region of the image. The diffraction spike is also in this region, but was suppressed using correction techniques described in the IRAC Handbook.

\subsection{IRS}

Spectra of SW1 covering 7.4-38.0 $\mu$m were obtained with the IRS operating in the Long-Low order 1, Long-Low order 2, and Short-Low order 1: LL1, 19.5-38.0 $\mu$m, LL2, 14.0-21.3 $\mu$m, and SL1, 7.4-14.5 $\mu$m. The angular size of the slits on the sky are 11$''$ $\times$ 170$''$ for the LL modes and 3$''$.6 $\times$ 57$''$ for the SL mode. The exposure time for each of the observations can be seen in Table 1. A detailed summary of IRS and spectral reduction and analysis techniques can be found in the IRS Instrument Handbook (IRS instrument Team \& Science User Support Team 2011). The Standard Staring Mode of operation was used, which included a nodding of the telescope.

\subsection{MIPS}

The Multiband Imaging Photometer for {\it Spitzer} (MIPS) is an infrared imager with 24.0, 70.0, and 160.0 $\mu$m bands. A detailed summary of MIPS can be found in the MIPS handbook available through the {\it Spitzer} Heritage Archive (MIPS Instrument and Instrument Teams 2011). SW1 was observed in all three bands, but only recovered in the 24.0 $\mu$m and 70.0 $\mu$m images. Observations with MIPS involved scanning a large region around SW1, which resulted in most pixels in the mosaic having an $\sim$70 second exposure for the 24.0 $\mu$m and $\sim$40 second exposure for the 70.0 $\mu$m image. These images, after MOPEX mosaicking and calibration, can be seen in Figure 2, each having an 8.0$'$$\times$8.0$'$ FOV. Figure 3(a) shows the full 24 $\mu$m mosaic of SW1 along with intensity contours showing the coma shape and structure. This figure can be related to Stansberry et al. (2004) Figure 1, but we have oriented the images with Equatorial North up as opposed to Ecliptic North. Also, Figure 3(b) shows the 24 $\mu$m image with a 1/$\rho$ profile removed similar to Stansberry et al. (2004) Figure 2 {\it lower left} panel ($\rho$ is the skyplane projected cometocentric distance).

\begin{figure}
        \centering
        \begin{subfigure}[b]{0.25\textwidth}
                \includegraphics[scale = 0.27]{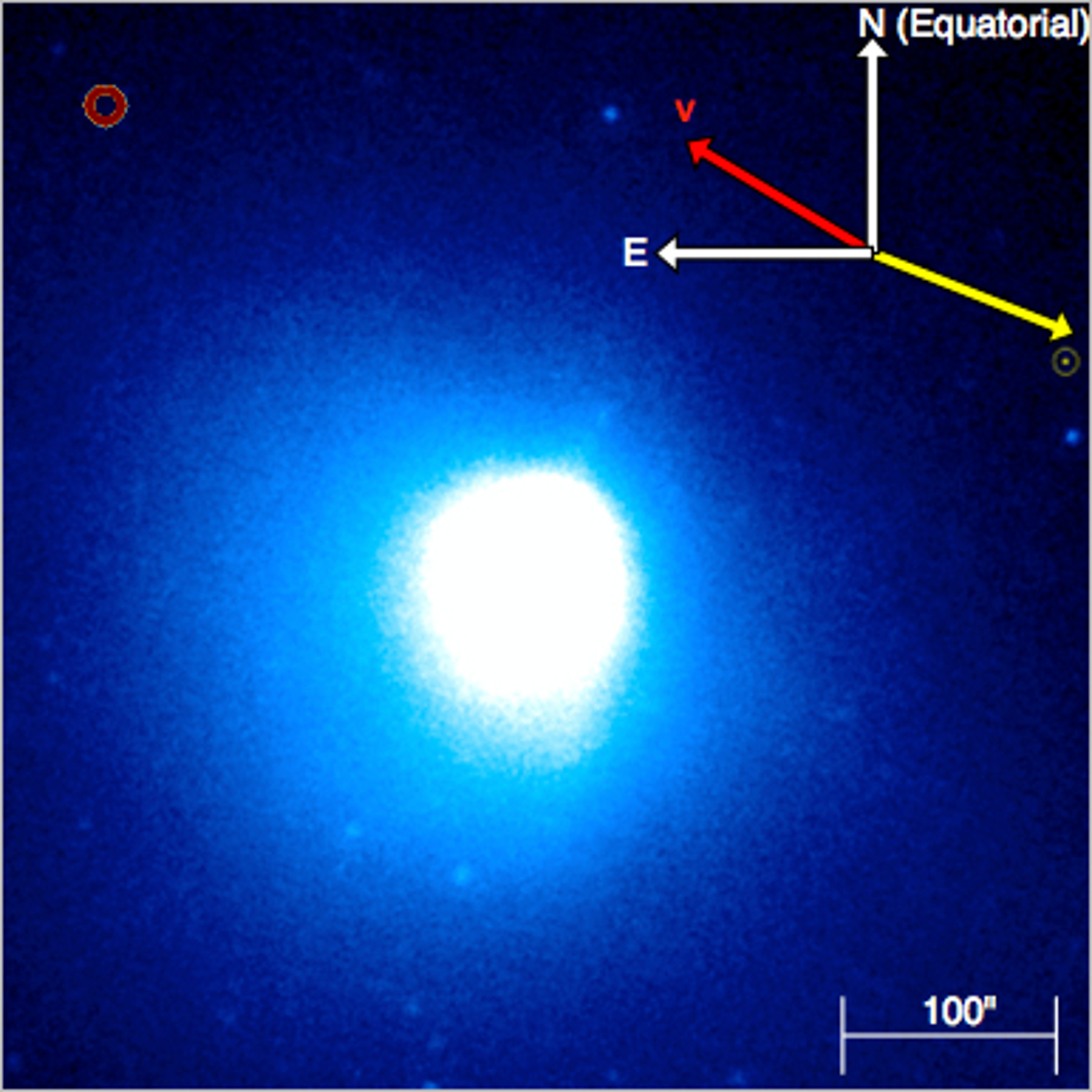}
                \caption{}
                \label{fig:comet}
        \end{subfigure}%
        ~~~~~~~~~~~~~~~~~~~~~~~~~~        
        \begin{subfigure}[b]{0.25\textwidth}
                \includegraphics[scale = 0.27]{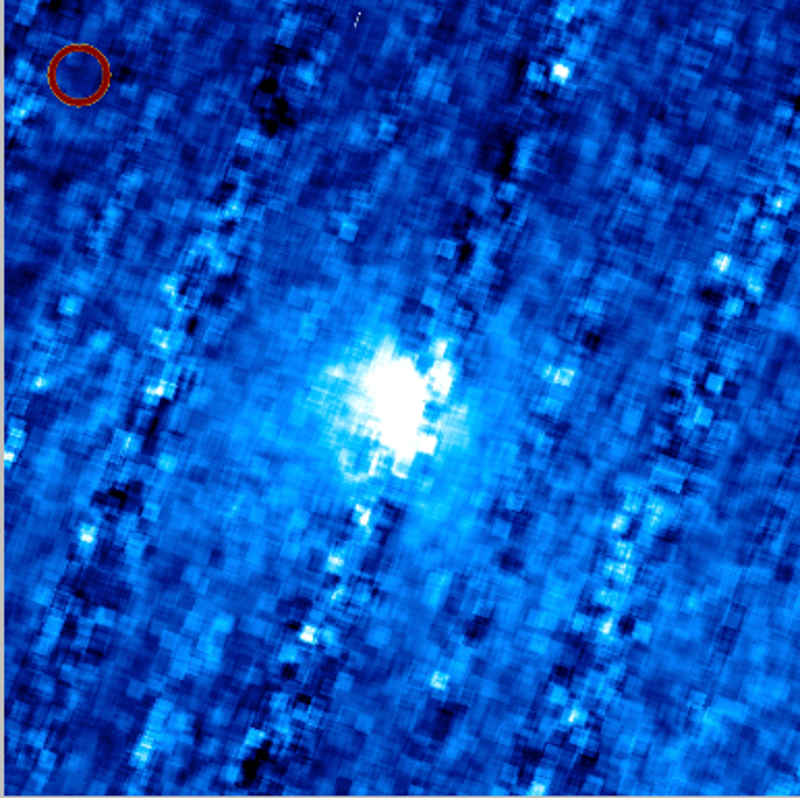}
                \caption{}
                \label{fig:tiger}
        \end{subfigure}
        \caption{(a) 24.0 $\mu$m and (b) 70.0 $\mu$m mosaicked and calibrated image centered on SW1.  Each has a FOV of  8.0$'$ $\times$ 8.0$'$ and equatorial north is up and east is to the left. The solar direction is identified by the yellow arrow and SW1's projected heliocentric velocity by the red arrow. The scale bar at the bottom right of (a) shows a distance of 100$''$ and the circle in the top left of the images represents the size of the PSF's first minimum.}
\end{figure}

MIPS mosaicked images are notorious for image artifacts in the final MOPEX generated mosaic. Each of these image artifacts are described in detail in the MIPS handbook, along with correction algorithms to mitigate their impact on the final mosaic. Processing for the 24 $\mu$m image involved applying a median time filter per pixel, which removed the long lived dark latent artifacts caused by a $>$ 50 Jy source being imaged after the last array annealing, but before the SW1 observations. The 70 $\mu$m image suffered from artifacts resulting from slow response variations on the array leading to streaks and stim latents due to the stim flashes used to measure the slow response of the array. Two correction algorithms were implemented on the BCDs before mosaicking which improved the quality of the final image: high-pass median time filter per pixel and column median value subtraction. Traces of a ``jail-bar" artifact can still be seen in Figure 2 (b).

\begin{figure}

	\centering
	\begin{subfigure}[b]{0.5\textwidth}	
		\includegraphics[scale = 0.6]{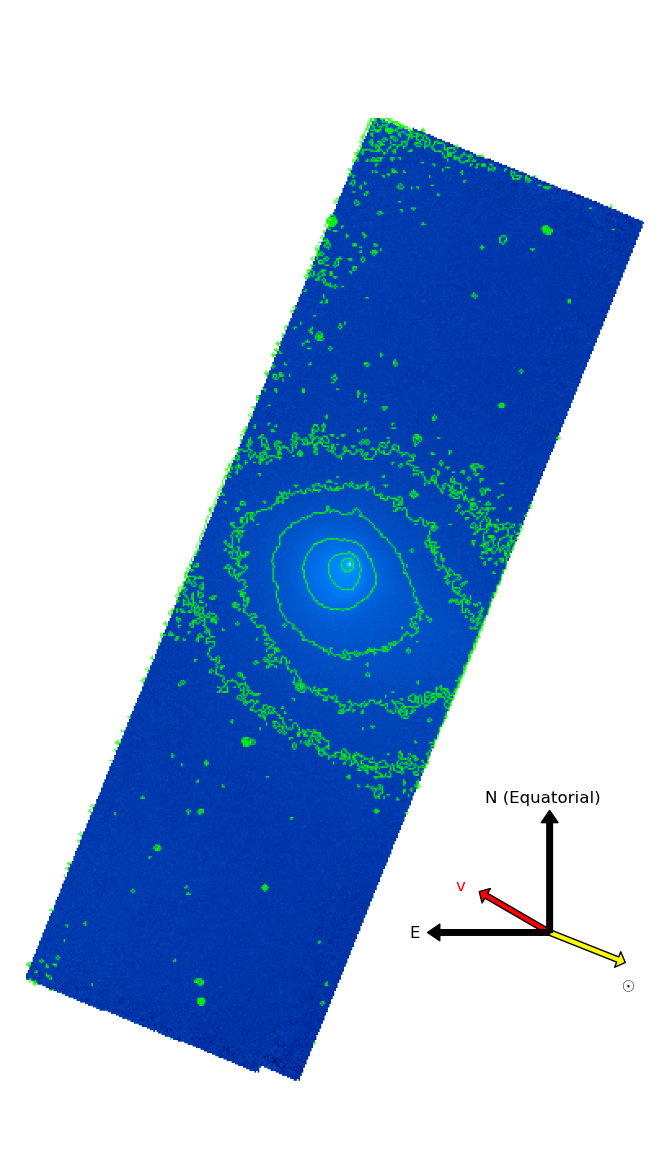}
		\caption{}
		\label{fig:comet}
	\end{subfigure}%
        ~~~~~~~~~~~~~
        \begin{subfigure}[b]{0.5\textwidth}
                \includegraphics[scale = 0.3]{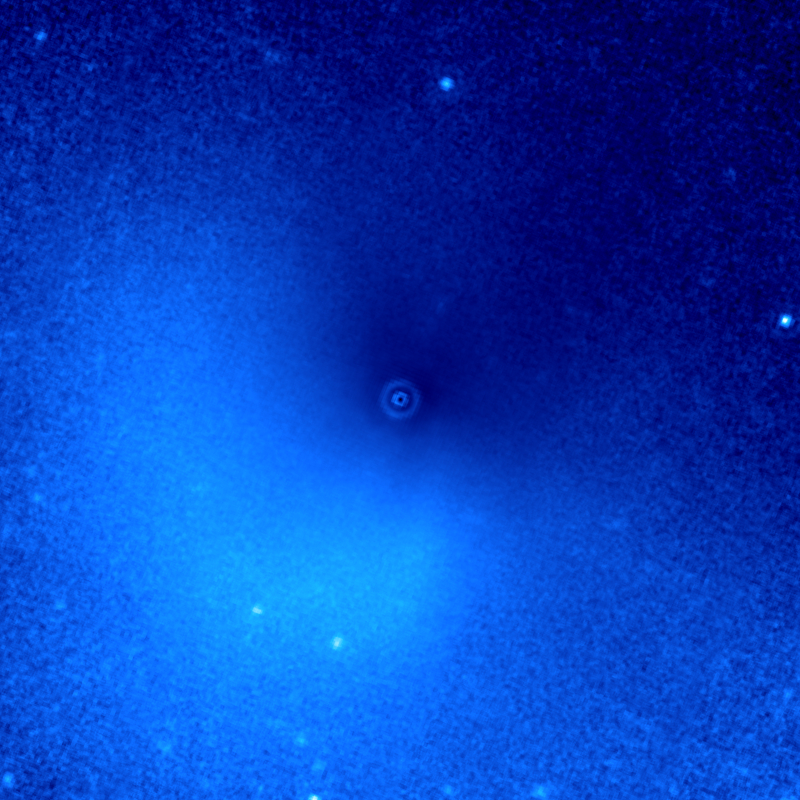}
                \caption{}
                \label{fig:tiger}
        \end{subfigure}	
\caption{(a) 24 $\mu$m MIPS mosaic of SW1 showing contours to highlight the coma and tail shape. The mosaicked image has dimensions of (17$' \times 52'$). The contour levels of the image are (0.2, 0.5, 1.0, 2.5, 5.0 and 10.0) MJy/sr. (b) 1/$\rho$ removed 24.0 $\mu$m image similar to Stansberry et al. (2004) Figure 2 ({\it lower left}) panel, which shows enhancement in the south-east region produced by the single jet observed in the 2004 analysis. It should be noted that the jet feature is due to solar radiation pressure and not from rotation as was previously believed. The scale and orientation of this image are the same as Figure 2(a). }

\end{figure}

\section{Nuclear Properties}

Measuring nuclear properties of active comets is hindered by the presence of emission from the gas and dust contained in the coma. For an accurate assessment of the thermal radiation emitted by SW1's nucleus the coma flux must be modeled and removed. This process is one of painstakingly and meticulously comparing the residuals of an image, with a model coma removed, to a point-spread function (PSF) of the optical system used for the observations. It should be pointed out that for most comet observations the nucleus is unresolvable in the image and its contribution to the image is a PSF. Thus, using a scaled PSF as a measure for the flux contribution is warranted. Our software was developed by our team and is based on the technique described by Lamy and Toth (1995) and used many times since then (e.g., Lamy et al. 2006, 2011; Fernandez et al. 1999, 2013; Kelley et al. 2013), in particular, to correctly predict the sizes of nuclei before their flyby by spacecraft (Lamy et al. 1998; Fernandez et al. 2003;  Lisse et al. 2009). In this section a description of the nucleus extractions techniques is discussed, the resulting nuclear spectral flux density measurements are presented, and nuclear properties extracted from thermal modeling are shown.

\subsection{Coma Removal}

The modeling and removal of the coma from a comet's image is not a straightforward procedure or ``one size fits all" technique that can be applied to comet observations. Each image must be analyzed to determine a modeling and removal method best suited for a successful point source extraction. Two modeling techniques were found to be necessary for the {\it Spitzer} images: a scaling of a system PSF for the 5.8 $\mu$m and 8.0 $\mu$m images and a more complex coma modeling routine for the 24.0 $\mu$m and 70.0 $\mu$m bands. Each method will be described in detail in the following paragraphs. Both have been tested with synthetic comet images to verify coma removal capabilities. 

For the 5.8 $\mu$m and 8.0 $\mu$m images, shown in Figure 1, there was not sufficient flux from the coma to produce a usable coma model. In addition, the removal of the diffraction spike from the images resulted in an artifact that interfered with any coma flux present. This method involved first finding the pixel location of the comet's centroid and choosing this to be the location of the nucleus. Next, a PSF was centered on this location and scaled to match the level of the comet's flux. Initially, a PSF generated from the program STINYTIM (Krist 2006) was used for this procedure, but it was found that this PSF did not represent well the structure observed in the comet's image or field stars also in the image. One of the field stars from each band, sufficiently distant from the diffraction spike, was used to represent the PSF. SW1 was moving slow enough during each image's exposure that the projected distance traveled was less that a pixel scale length, which was approximately 4800 km/pixel. Once the PSF was scaled it was subtracted from the comet image, resulting in what should be only flux from the coma and background. A best-fit PSF was taken to be one that minimized the standard deviation of the residual of pixels from a region centered on the comet's centroid. For the 5.8 $\mu$m image there was not much come and the diffraction spike also presented a problem so this was the best approach found to measure the nuclear flux.  This scaled PSF was then taken to represent the nucleus's contribution to the image flux and used for photometry measurements. Figure 4 shows radial profiles of the 5.8 $\mu$m comet image and scaled PSF for a selection of azimuthal angles. Figure 5 shows the 5.8 $\mu$m comet image, scaled PSF image, and residuals after PSF subtraction. Similarly, Figures 6 and 7 are for the 8.0 $\mu$m image.

\begin{figure}[h!]
\begin{centering}
\includegraphics[scale = 0.45]{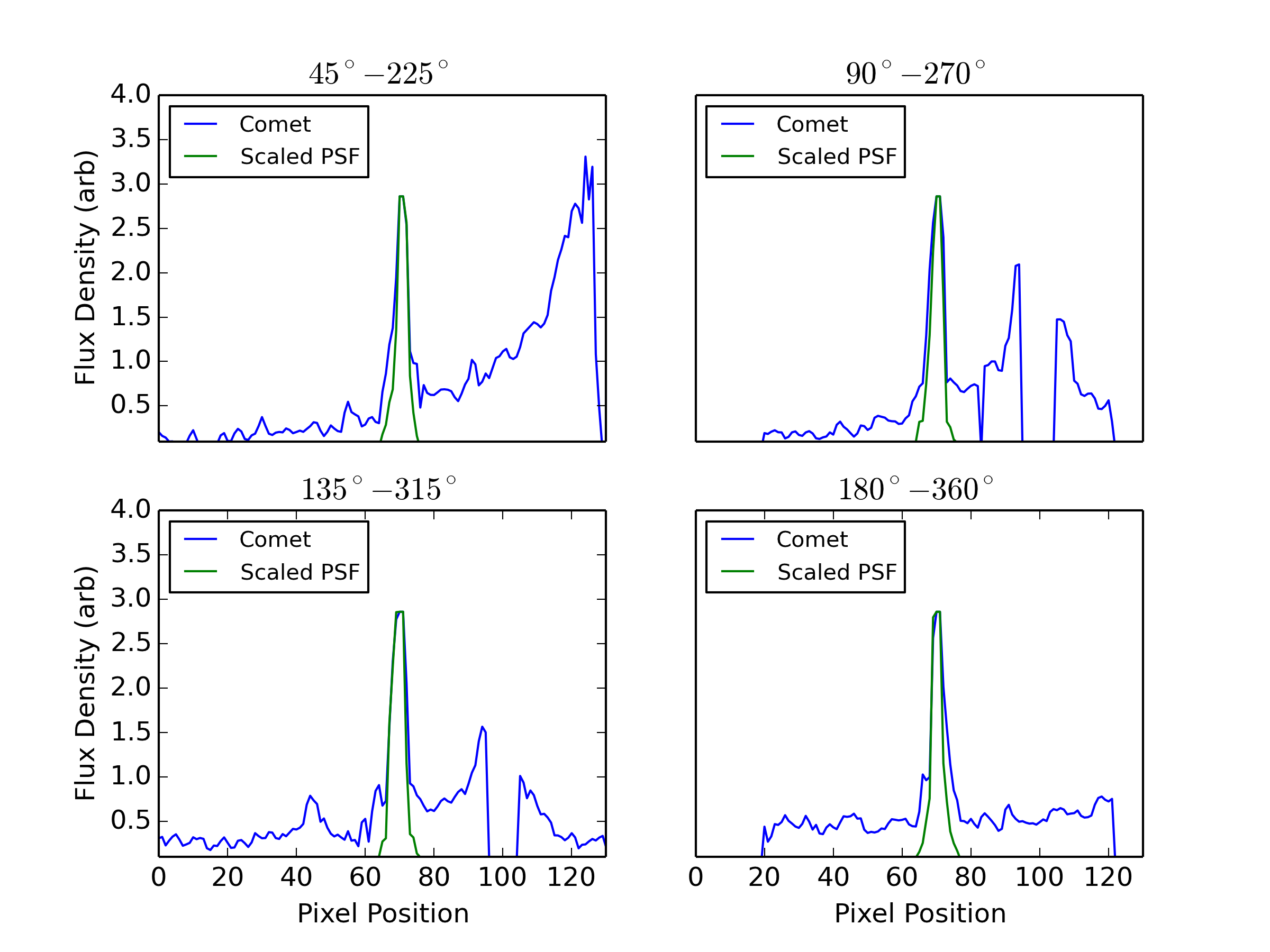}
\caption{Radial cross-section of the 5.8 $\mu$m image showing the scaled PSF. The angular descriptor indicates the position angle of the radial cross-section. The coma and artifact contributions to the images can be seen in the comet profiles. Note that there is not much coma in the core. }
\end{centering}
\end{figure}

\begin{figure}[h!]
	\centering
	\begin{subfigure}[b]{0.25\textwidth}
		                \includegraphics[scale = 1.2]{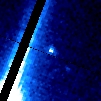}
		                \caption{}
	\end{subfigure}%
        ~
	\begin{subfigure}[b]{0.25\textwidth}
		                \includegraphics[scale = 1.2]{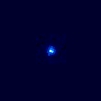}
		                \caption{}
	\end{subfigure}
	
	\begin{subfigure}[b]{0.25\textwidth}
		                \includegraphics[scale = 1.2]{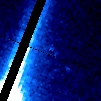}
		                \caption{}
	\end{subfigure}
	\caption{The (a) 5.8 $\mu$m comet image, (b) the scaled PSF, and (c) the residual after PSF subtraction.}
\end{figure}

\begin{figure}[h!]
\begin{centering}
\includegraphics[scale = 0.45]{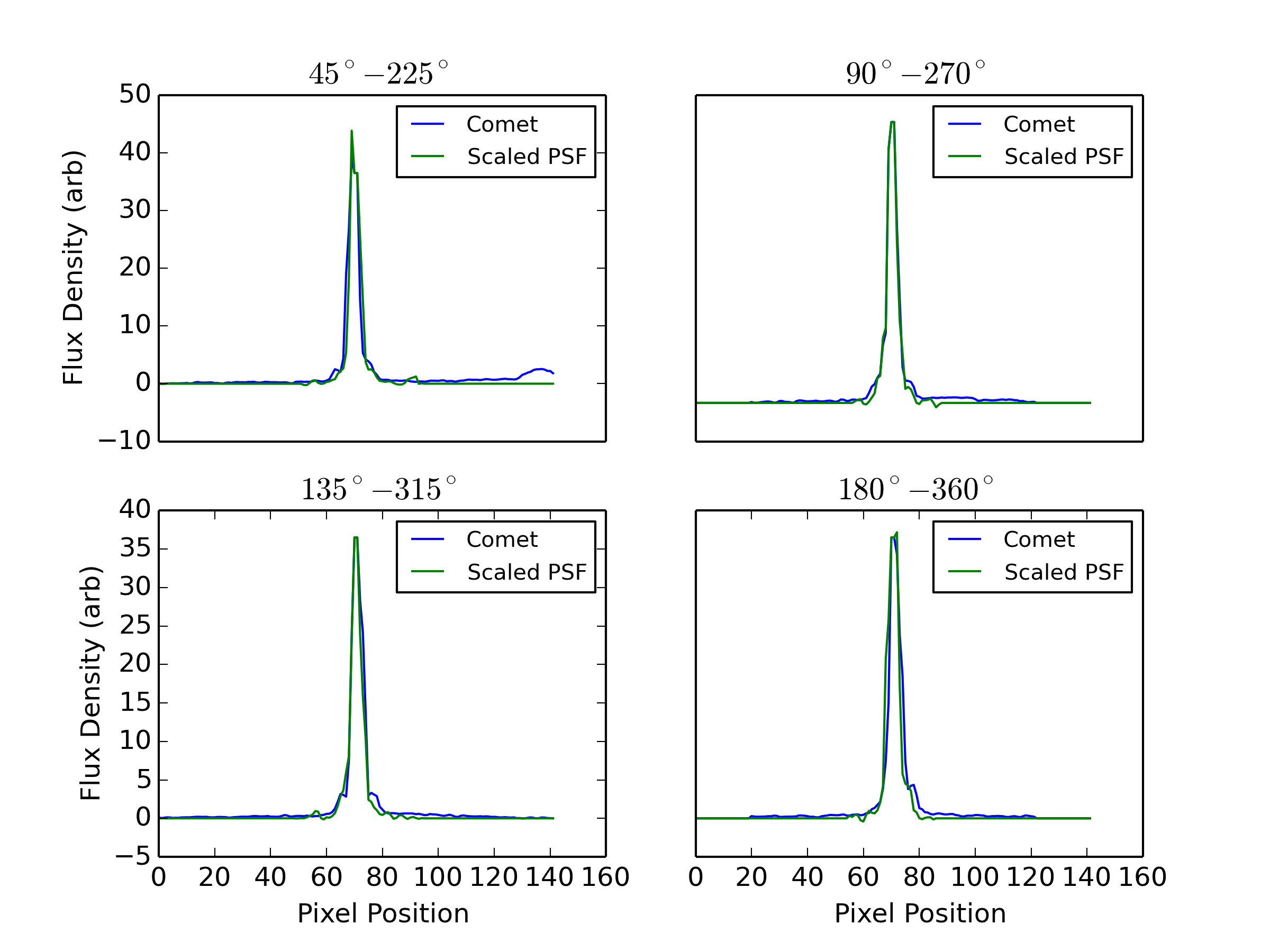}
\caption{Radial cross-section of the 8.0 $\mu$m image showing the scaled PSF. The angular descriptor indicates the position angle of the radial cross-section. Note there is not much come coma present in the core.}
\end{centering}
\end{figure}

\begin{figure}[h!]
	\centering
	\begin{subfigure}[b]{0.25\textwidth}
		                \includegraphics[scale = 1.2]{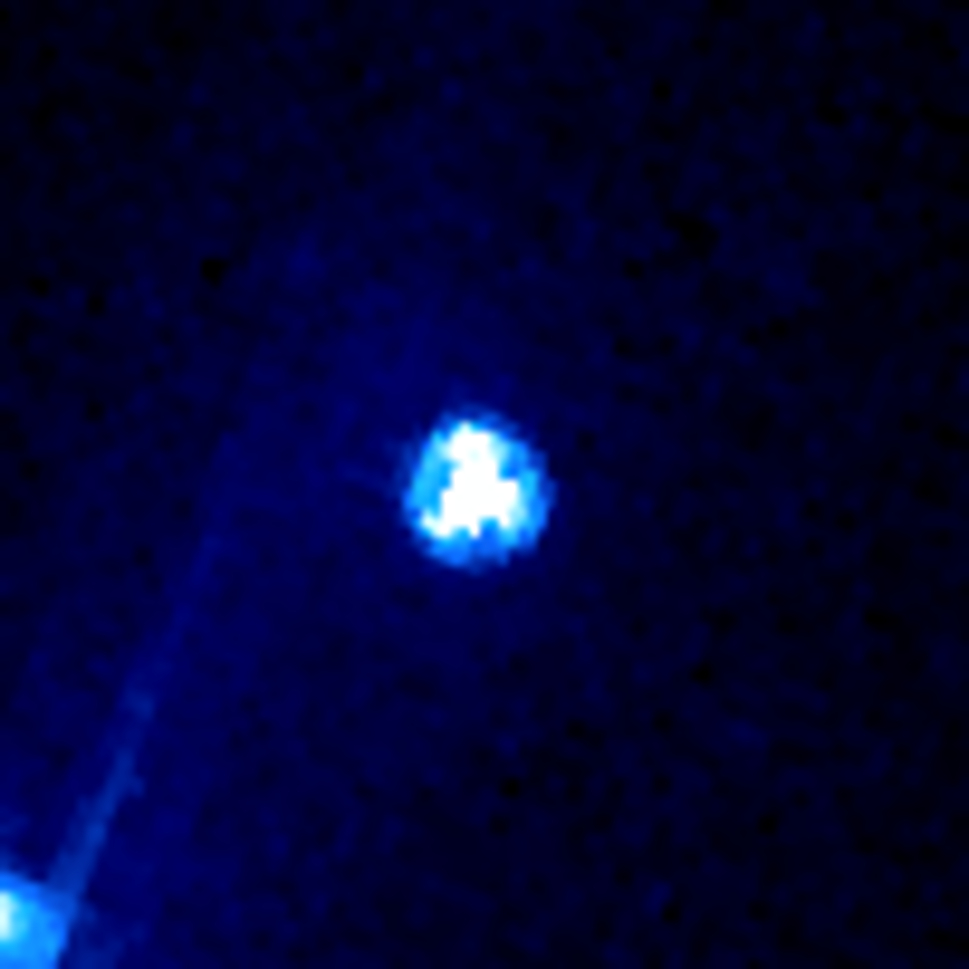}
		                \caption{}
	\end{subfigure}%
        ~
	\begin{subfigure}[b]{0.25\textwidth}
		                \includegraphics[scale = 1.2]{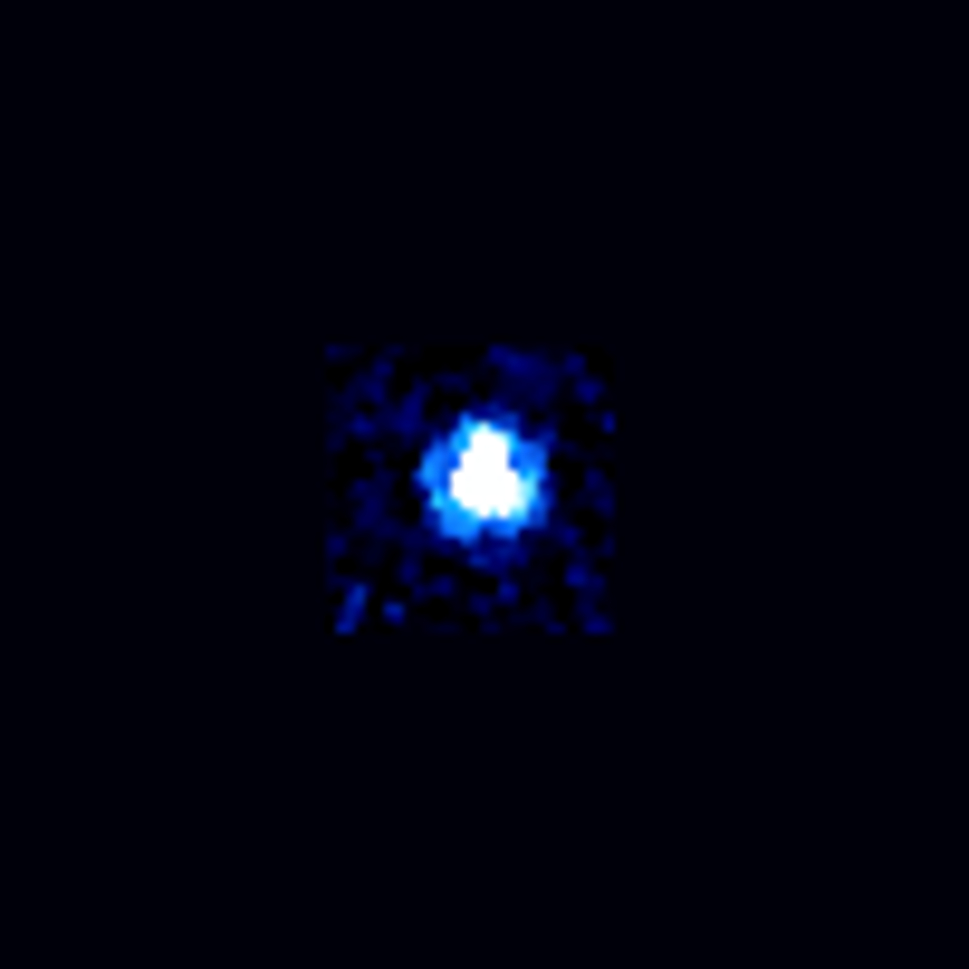}
		                \caption{}
	\end{subfigure}
	
	\begin{subfigure}[b]{0.25\textwidth}
		                \includegraphics[scale = 1.2]{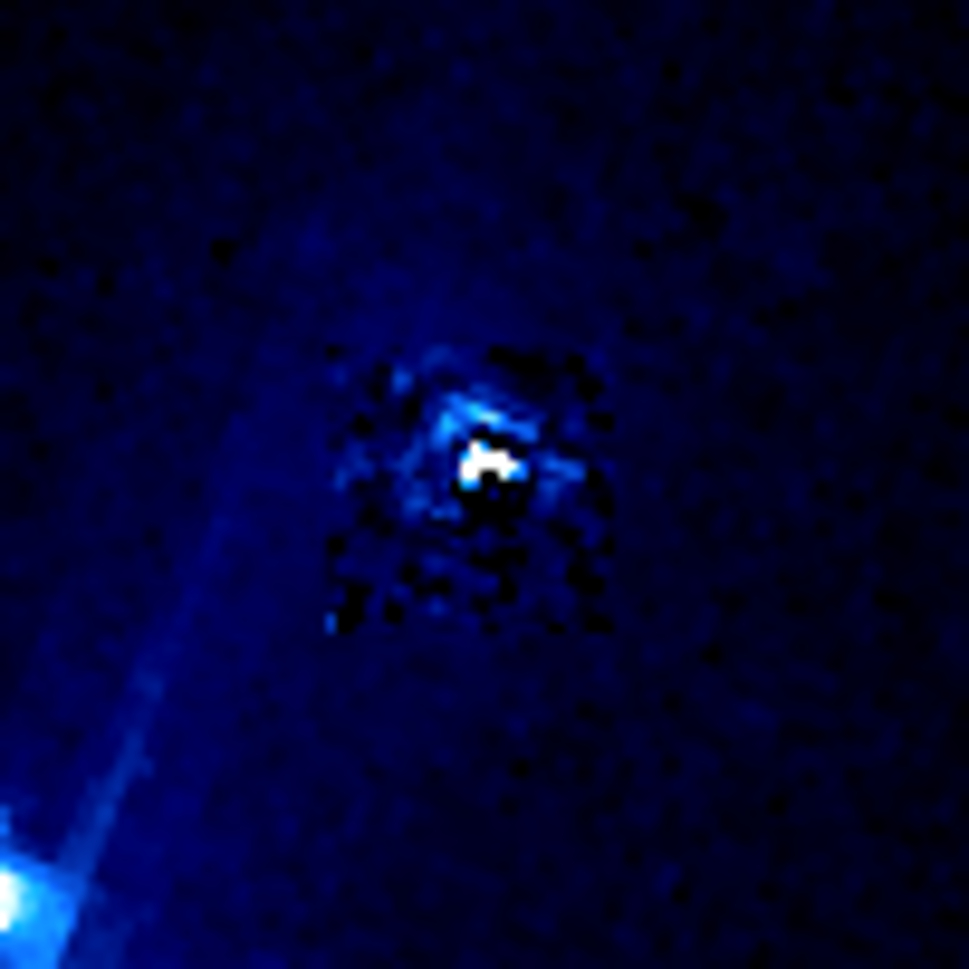}
		                \caption{}
	\end{subfigure}
	\caption{The (a) 8.0 $\mu$m comet image, (b) the scaled PSF, and (c) the residual after PSF subtraction. Notice in (c) there is an over subtraction to the south of the centroid and under subtraction to the north of the centroid. This can be attributed to asymmetry in the star generated PSF and SW1 nucleus image.}
\end{figure}

The 24.0 $\mu$m and 70.0 $\mu$m images had sufficient coma present for a more elaborate coma modeling routine to be implemented on these images (Lamy et al. 2004; Lisse et al. 2009; Fernandez 1999). The procedure for this method is to first take azimuthal pie sections of the comet image, centered on the comet's centroid. Radial profiles of each pie section are generated. The angular width of the sections was a variable for the procedure and was selected independently for each image. The S/N of the image determined the minimum angular spread usable, with smaller angles being desired for higher fidelity in the model coma. Again, since the nucleus is unresolvable to {\it Spitzer} at this distance it will only contribute flux in the region of the PSF. In theory the nucleus should only image as a PSF on the image plane. Thus, any flux present outside of the region of the PSF should be contributions from the coma and background. An example of these radial profiles from both the 24.0 $\mu$m and 70.0 $\mu$m image and their corresponding PSFs can be seen in Figures 8 and 9. If regions radially distant from the nucleus (regions outside the dominance of the PSF) are chosen, a coma model of the form $A(\theta)/r^{n(\theta)}$ can be fit to the profiles, generating a synthetic coma. The parameter $\theta$ is the position angle (PA) of the pie section used for the radial profile. This procedure was implemented on each of the azimuthal pie section producing the coma model shown in Figure 10(b) for the 24.0 $\mu$m band and Figure 11(b) for the 70.0 $\mu$m band. The coma model is then subtracted from the comet image yielding the PSF's contribution to the image. A comparison of the residual PSF and an STINYTIM-generated PSF is shown in Figures 10,11(c) and (d). Close inspection of the two shows a high degree of similarity (the first order bright fringe as an example), giving a level of validation to our modeling technique.

\begin{figure}
\begin{centering}
\includegraphics[scale = 0.45]{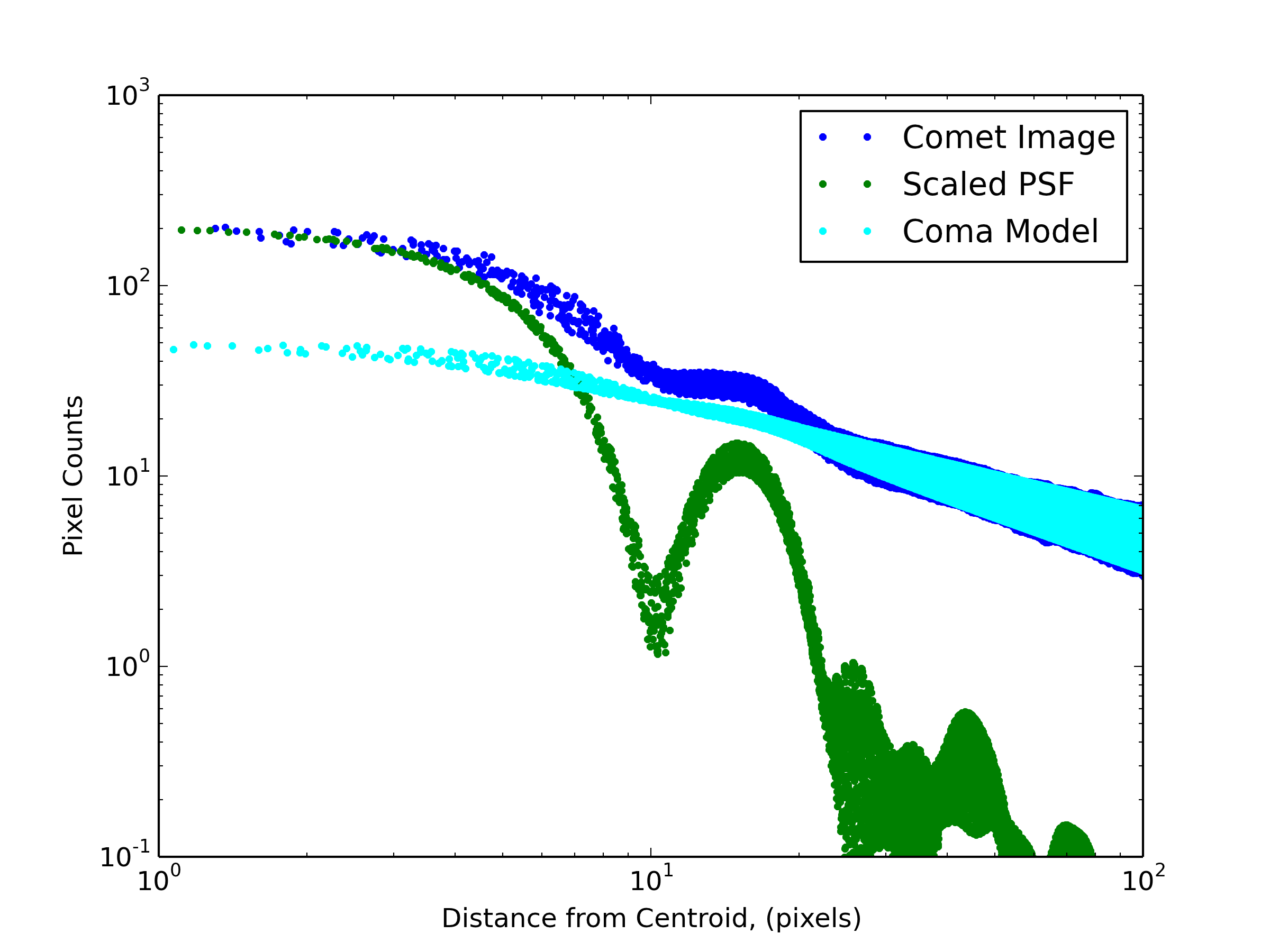}
\caption{Shown are radial profiles of the 24.0 $\mu$m comet image, STINYTIM PSF, and synthetic coma model. The PSF effectively goes to zero at a certain radial position which can be seen to be around 30 pixels. Comet flux beyond this region was used for the fitting procedure. Extrapolation of the coma model to the center shows the excess contribution to the comet image attributed to the nucleus.}
\end{centering}
\end{figure}

\begin{figure}
\begin{centering}
\includegraphics[scale = 0.45]{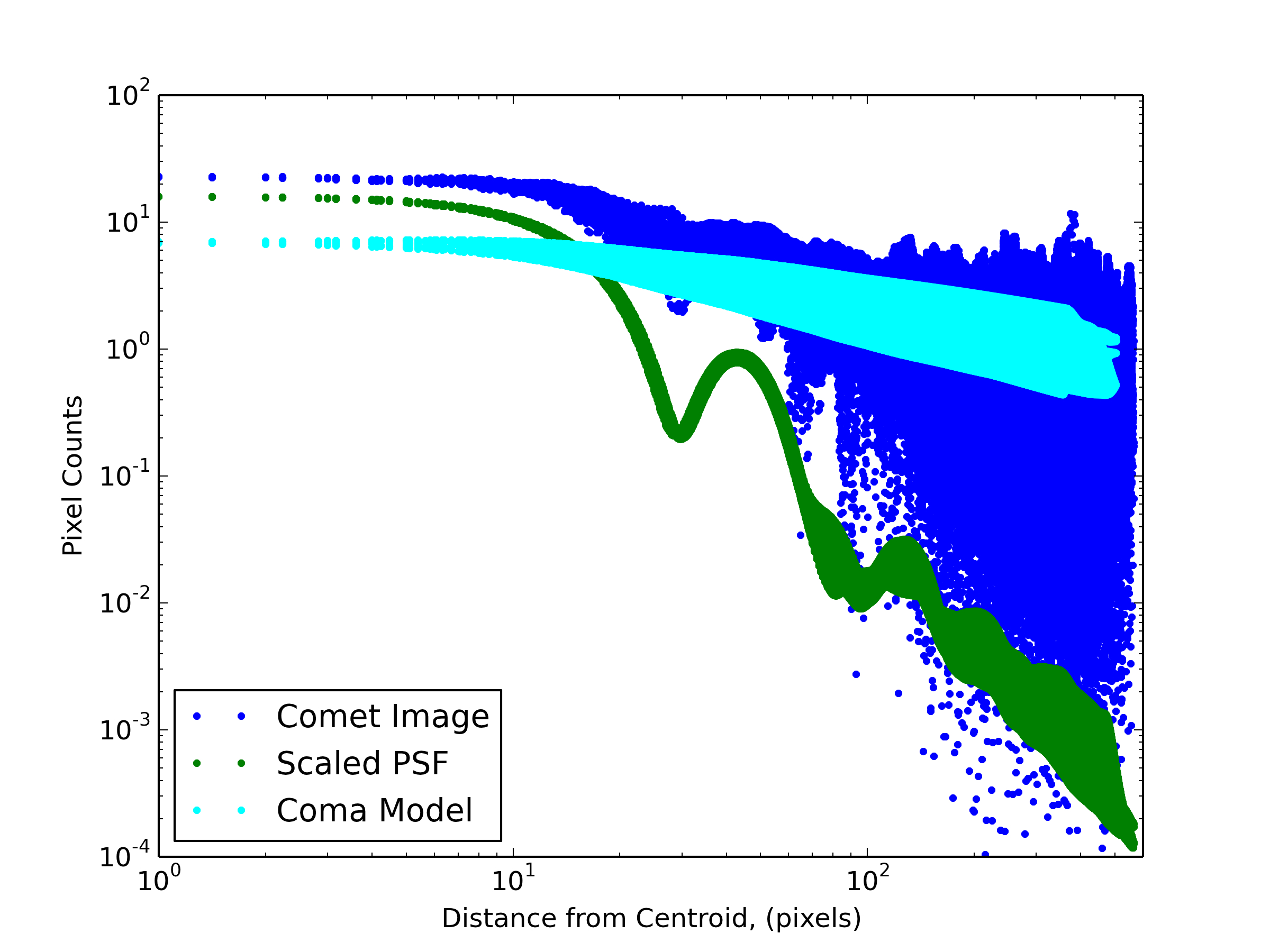}
\caption{Shown are radial profiles of the 70.0 $\mu$m comet image, STINYTIM PSF, and synthetic coma model, similar to the 24.0 $\mu$m profiles in Figure 8.}
\end{centering}
\end{figure}

\begin{figure}
        \centering
        \begin{subfigure}[b]{0.25\textwidth}
                \includegraphics[scale = 0.15]{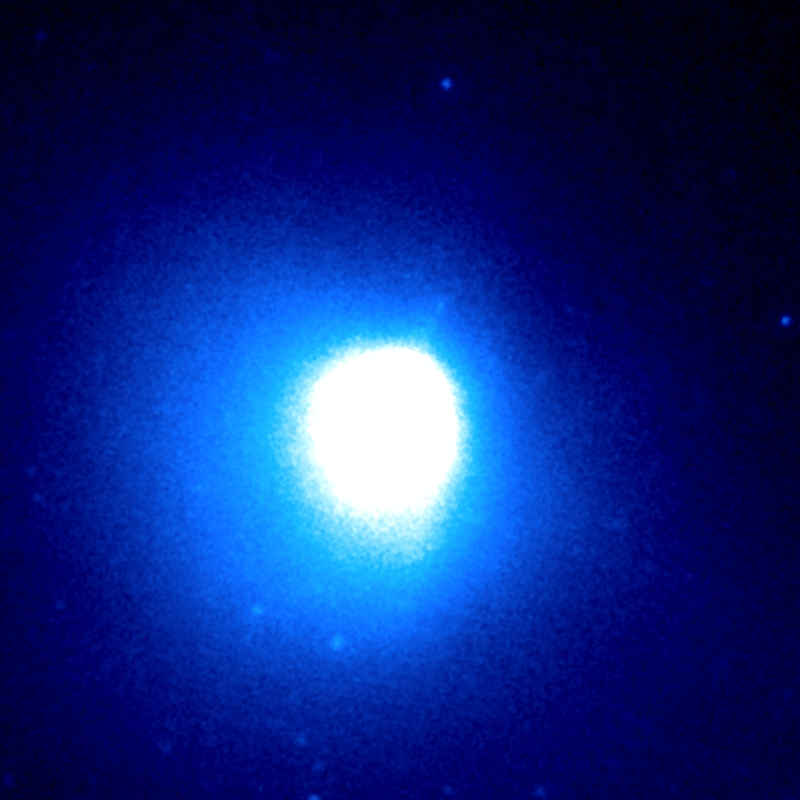}
                \caption{}
                \label{fig:comet}
        \end{subfigure}%
        ~
        \begin{subfigure}[b]{0.25\textwidth}
                \includegraphics[scale = 0.15]{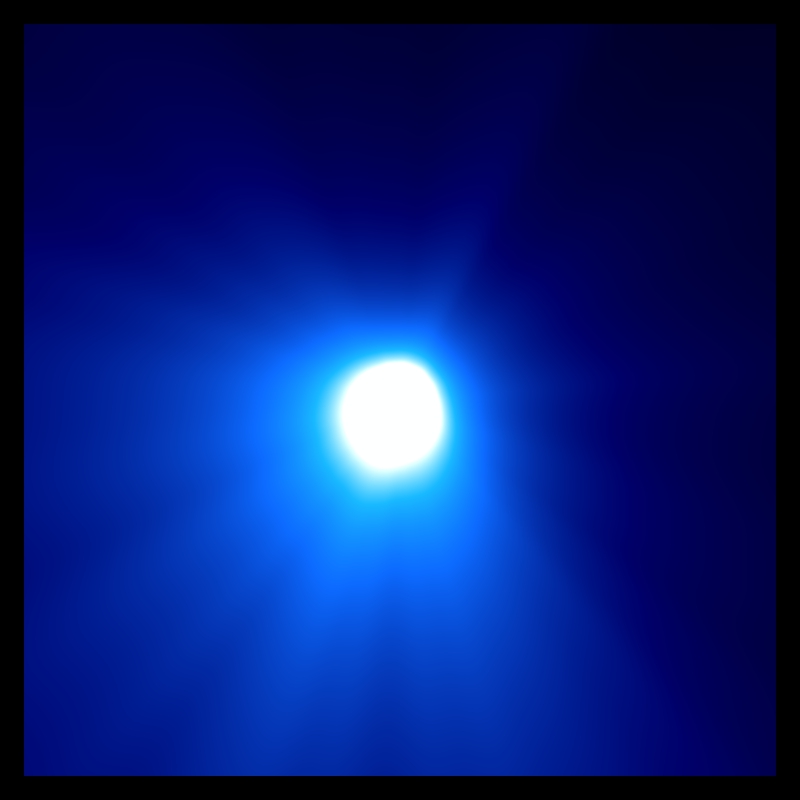}
                \caption{}
                \label{fig:tiger}
        \end{subfigure}
        ~ 
        \begin{subfigure}[b]{0.25\textwidth}
                \includegraphics[scale = 0.15]{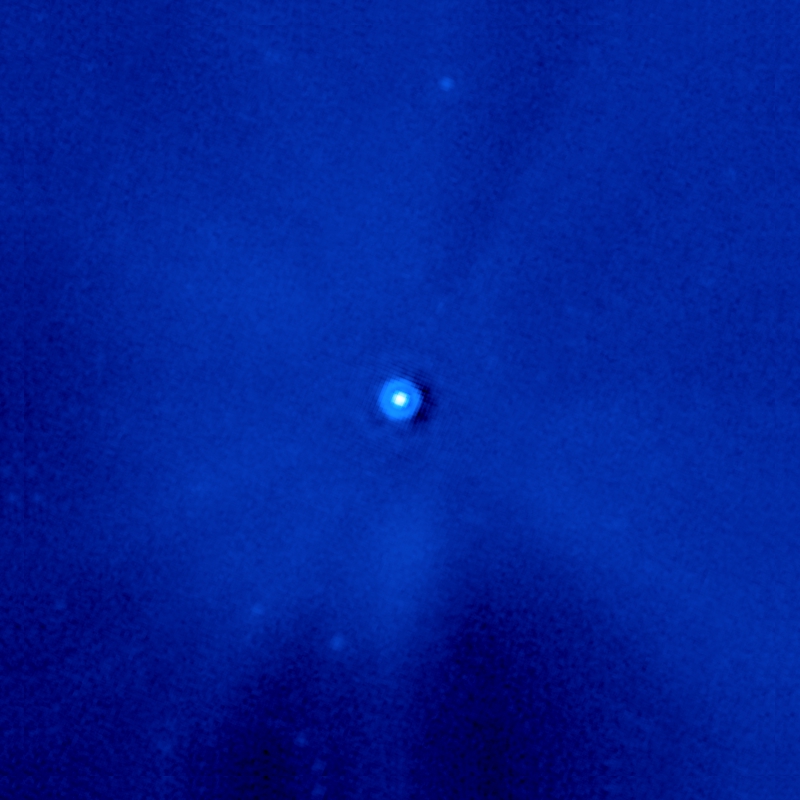}
                \caption{}
                \label{fig:diff}
        \end{subfigure}%
          ~ 
        \begin{subfigure}[b]{0.25\textwidth}
                \includegraphics[scale = 0.15]{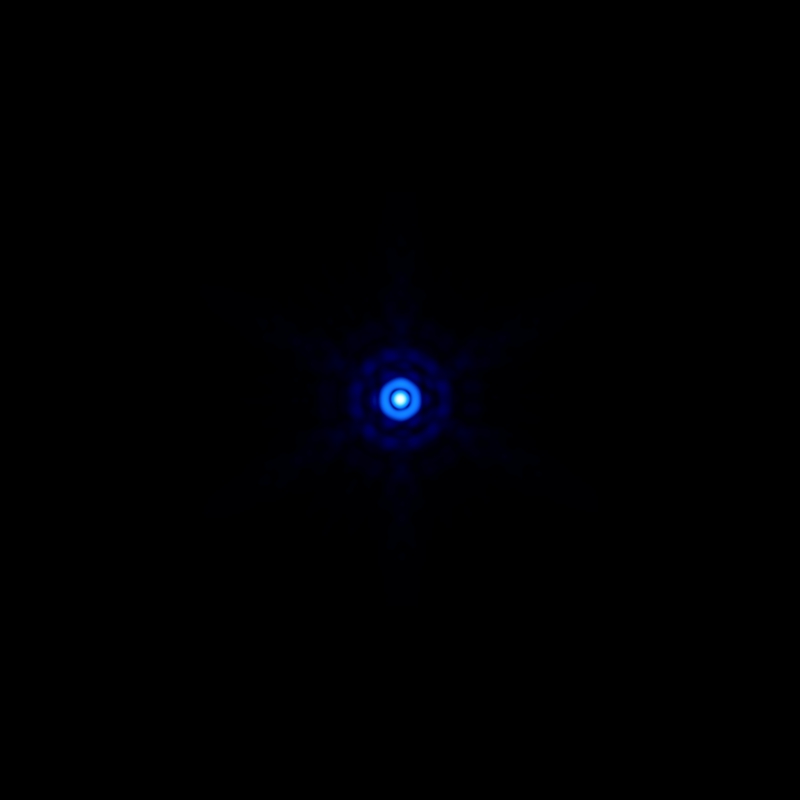}
                \caption{}
                \label{fig: PSF}
        \end{subfigure}
        \caption{(a) 24 $\mu$m comet image, (b) the synthetic coma model, (c) the residual after model coma removal, and (d) the STINYTIM 24 $\mu$m PSF. Notice the asymmetry seen in the comet image is modeled well in the coma. Also, the residual nucleus contribution has many features similar to the optical system PSF, indicating a good coma removal.}
\end{figure}

\begin{figure}
        \centering
        \begin{subfigure}[b]{0.25\textwidth}
                \includegraphics[scale = 0.15]{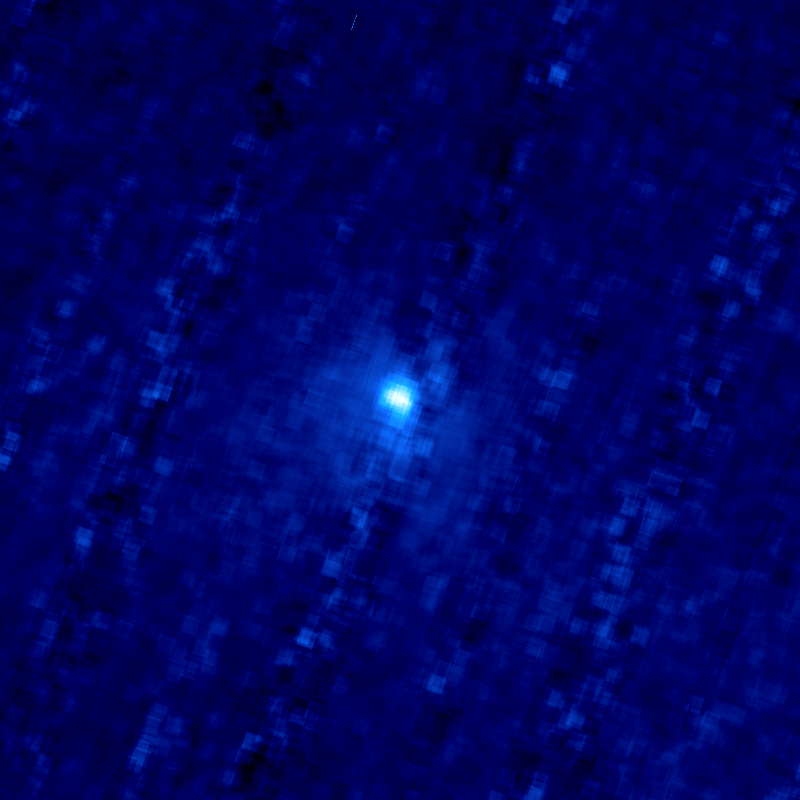}
                \caption{}
                \label{fig:comet}
        \end{subfigure}%
        ~
        \begin{subfigure}[b]{0.25\textwidth}
                \includegraphics[scale = 0.15]{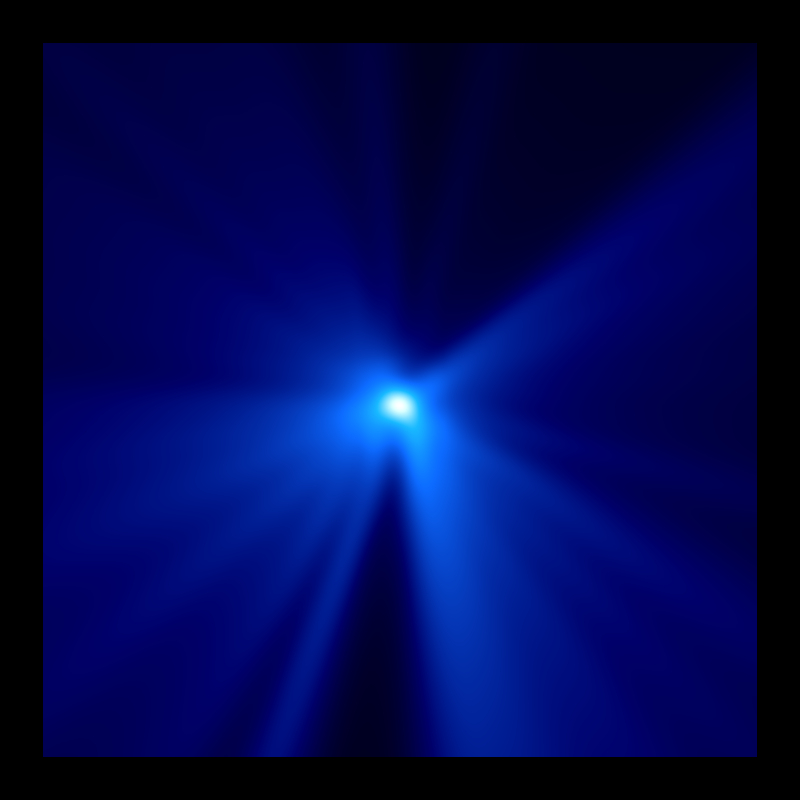}
                \caption{}
                \label{fig:tiger}
        \end{subfigure}
        ~ 
        \begin{subfigure}[b]{0.25\textwidth}
                \includegraphics[scale = 0.15]{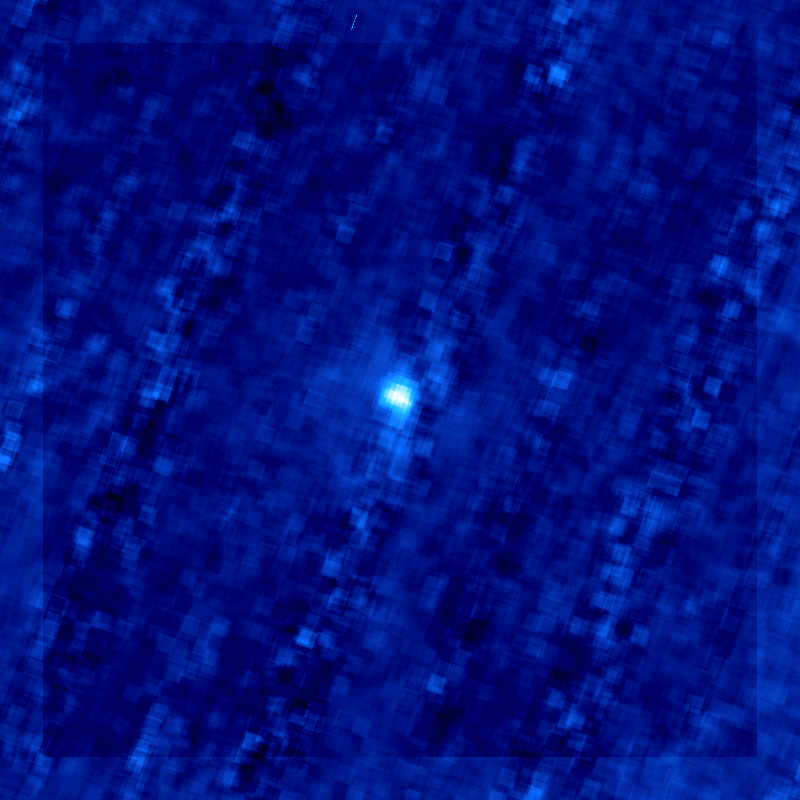}
                \caption{}
                \label{fig:diff}
        \end{subfigure}%
          ~ 
        \begin{subfigure}[b]{0.25\textwidth}
                \includegraphics[scale = 0.15]{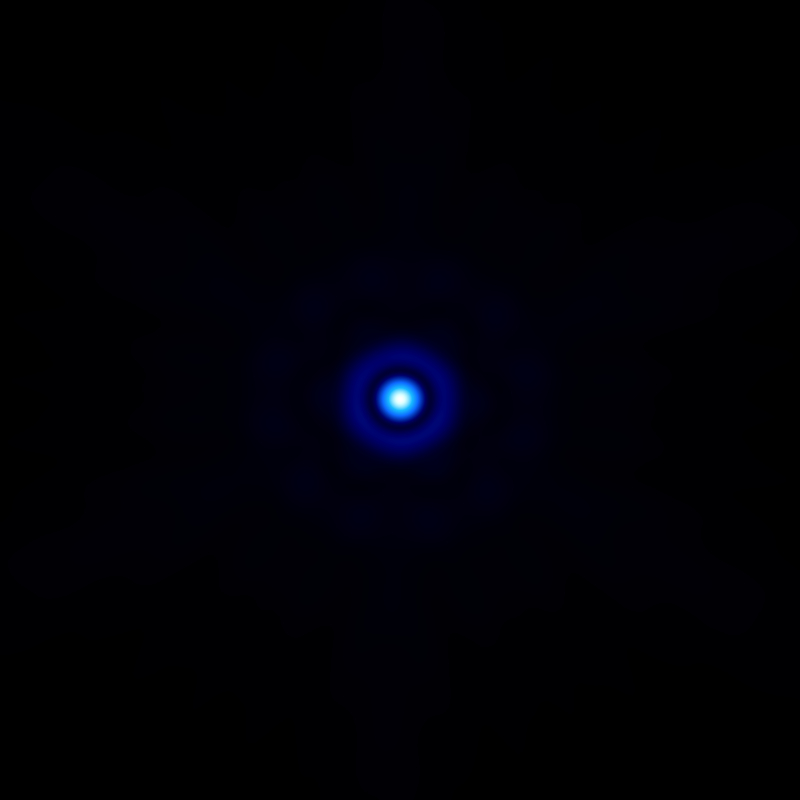}
                \caption{}
                \label{fig: PSF}
        \end{subfigure}
        \caption{(a) 70 $\mu$m comet image, (b) the synthetic coma model, (c) the residual after model coma removal, and (d) the STINYTIM 70 $\mu$m PSF.}
\end{figure}

\subsection{Nuclear Photometry}

Once the coma was successfully removed from the images, photometry was used to measure the spectral flux density in each of the four observational bands. Since the coma removal process resulted in a scaled PSF representing the contribution to the spectral flux density from the nucleus, no aperture correction was needed for the photometry. Table 2 shows the photometry results after the application of color corrections. Uncertainties in the spectral flux density measurements are derived from analysis of the distribution of the flux measurements from the many coma models for each band. The procedure to arrive at the best-fit coma model involved varying parameter space of the model (centroid pixel location, angular size of the azimuthal pie section, size in pixels of the radial region used for fitting, center position of the radial bin of pixels) and observing how this minimized the difference between the residuals after coma removal for the case of the 24.0 $\mu$m and 70.0 $\mu$m and the standard deviation of the central pixels for the 5.8 $\mu$m and 8.0 $\mu$m images. Each of these coma models was stored and the resulting residuals from each model was measured for the nucleus' contribution to the spectral flux density of the image. The distribution of the flux measurements from the many coma models that we tried for each band was used to derive uncertainties for the photometry measurements. We varied the parameter space of the model (centroid pixel location, angular size of the azimuthal pie section, and radial region used for coma slope fitting) so as to determine how a model's answer for the nucleus photometry depended on these input parameter assumptions. The scatter in the nucleus photometry was used to derive the uncertainties we list for the nucleus in Table 2.

With the nuclear photometry measured, the spectral flux density measurements for a $9''.0$ radius aperture of the coma's contributions were found for each of the IRAC and MIPS bands. These values can also be seen in Table 2. These values were found by subtracting the nuclear flux values from each image and then performing aperture photometry on the residual coma flux. Coma photometry is used to help tie together the orders that make up the IRS spectrum, and will be explained further in Section 4. The overall comet (coma + nucleus) photometry is well constrained, and the error bar for the coma is driven almost entirely by the error bar on the nucleus measurement.

\begin{center}
\begin{table}[h] 
\caption{}					
\resizebox{1.0\textwidth}{!}{

	\begin{tabular}{ l  c c c c c }
		\multicolumn{6}{ c }{Nuclear Photometry Measurements}\\
		\hline \hline
		Band	 ($\mu$m)		&	&   Coma Spectral Flux Density$^{\textrm{a}}$ (mJy) &	&	&	Nucleus Spectral Flux Density$^{\textrm{a}}$ (mJy)		\\
		                                  &      &   $9''.0$ Radius Aperture	           &    &     &              								\\ \hline
		5.6				& 	&   (1.51$^{+0.07}_{-0.1}$)$^{\textrm{b}}$			   &	 &	&	0.32$^{+0.07}_{-0.1}$	\\
		8.0				&	&   1.8 $\pm$ 1.7				   &	 &	&	4.6 $\pm$ 1.7			\\
		24.0				&	&   176.4 $\pm$ 13.4				   &	 &	&	199.8 $\pm$ 13.4		\\
		70.0				&	&   39.0 $\pm$ 28.3				   &	 &	&	175.4 $\pm$ 28.3		\\ \hline

	\end{tabular}

}
\caption*{
\resizebox{1.0\textwidth}{!}{
	\begin{tabular}{l}
		$^{\textrm{a}}$ Color corrections for spectral flux density measurements were calculated from methods described in the IRAC and MIPS handbooks.  \\
		$^{\textrm{b}}$ The 5.8 $\mu$m coma photometry measurement is most likely contaminated by presence of the diffraction spike, leading to the higher\\
		 measurement than the 5.8 $\mu$m nucleus photometry measurement.   \\

	\end{tabular}
}

}

\end{table}
\end{center}

\subsection{Thermal modeling}

The Near Earth Asteroid Thermal Model (NEATM, Harris 1998) is an improvement of the Standard Thermal Model (STM) in its applicability to many small bodies since it treats phase darkening in probably a more realistic way and since it lets the beaming parameter float as a free parameter. This model has been applied to many minor bodies (Mainzer et al. 2011; Delbo et al. 2011) and was applied to the {\it Spitzer} photometry measurements, returning values for the effective radius ($R$), IR beaming parameter ($\eta$), and geometric albedo ($p_{5.8}$). The IR beaming parameter is used as  a fitting parameter in  the NEATM, with a value of $\eta = 1$ applying for a perfectly memory-less, spherical surface with modest topography. Values of $\eta > 1.0$ imply lower surface temperatures than what would be observed for the ideal situation possibly caused by thermal communications between the day and night sides of the object. Topography and surface roughness can be inferred from values of $\eta < 1.0$. This is due to the observed surface temperature being higher than what would be observed from strictly applying the STM. Excess thermal flux can be thought to arise from emission from bowl shaped craters, where emission from the sides of the crater is absorbed at the bottom of the crater. This additional flux results in an increase in the equilibrium surface temperature when compared to the STM temperature. The beaming parameter is a way to adjust the equilibrium surface temperature distribution of the object and have it match the color temperature of the object's observations.

	The 5.8 $\mu$m photometry value is too high to be explained by thermal emission alone. Inclusion of a reflected light component, i.e. a scaled solar spectrum, into our model however explained the 5.8 $\mu$m flux. This is what would have let us constrain the V-band albedo p, in principle, although it turned out that the uncertainties were too great for a meaningful result for p. Our best fit model (using $\chi^2$ minimization), making use of both thermal and reflected components, yielded $R$ = 30.2 $^{-2.9}_{+3.7}$ km, $\eta = 0.99$ $^{-0.19}_{+0.26}$, and $p_{5.8} = 0.5 \pm 0.5$. Figure 12 shows the measured spectral flux density values, NEATM + Reflected curves, and best-fit spectral flux density values. Other parameters necessary for the thermal modeling are: bolometric bond albedo $A$ = 0.012 (assuming a visible-wavelenth geometrical albedo $p = 0.04$ and phase integral relation $q = 0.290 + 0.684$G, Harris and Lagerros 2002), emissivity $\epsilon = 0.95$, and slope parameter $G = 0.05$ (the same assumption made by Fern\'andes et al. 2013 in their {\it Spitzer} survey of cometary nuclei).

\begin{figure}
\begin{centering}
\includegraphics[scale = 0.3]{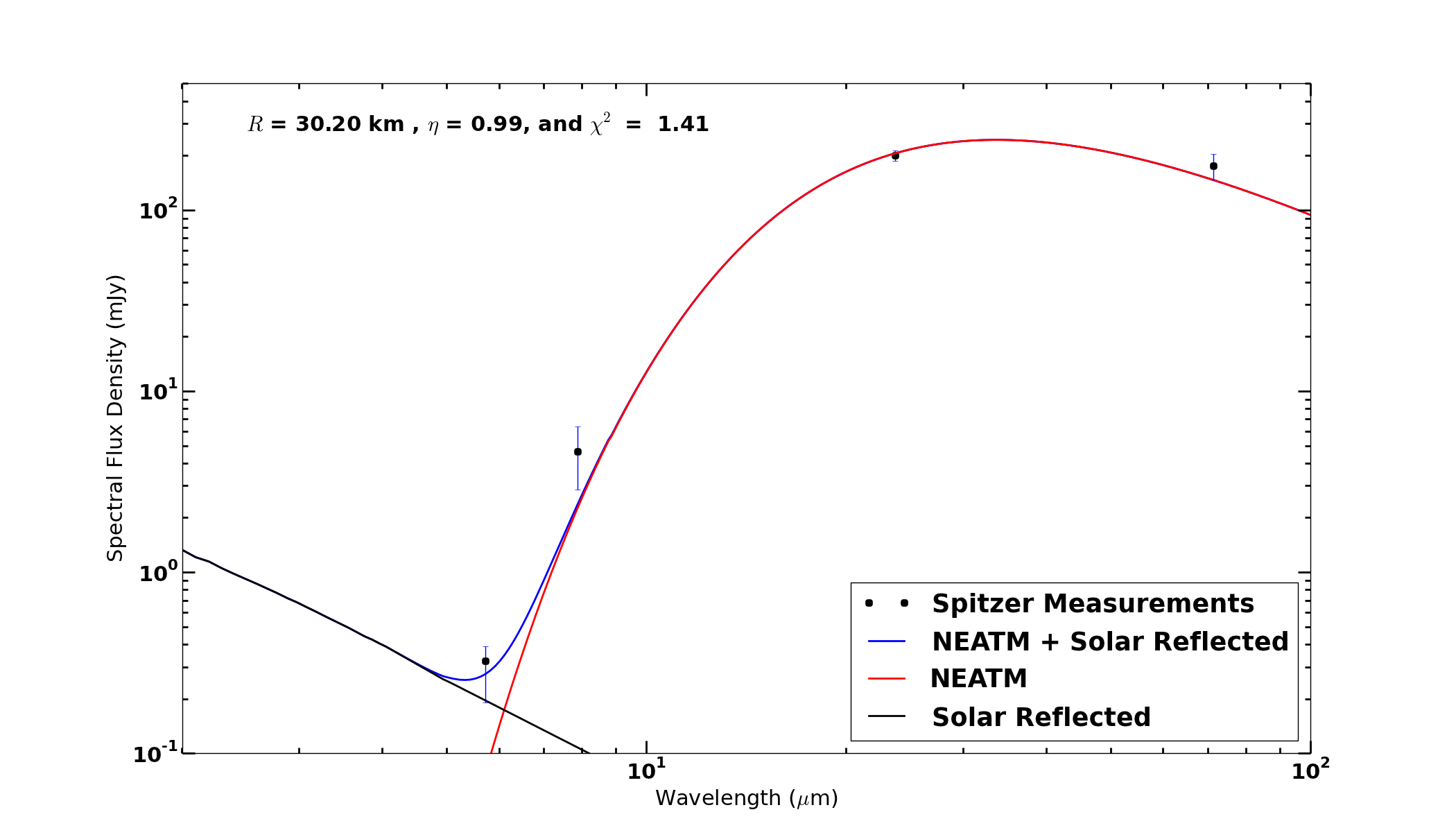}
\caption{Shown are the photometry measurements from {\it Spitzer} observations. Curves shown are of the NEATM fit to the measurements and reflected solar flux. 1-$\sigma$ error bars are included on the spectral flux density measurements.}
\end{centering}
\end{figure}

\section{Dust Compositions}

\subsection{Extended Source Calibration and Spectral Extraction}

IRS spectra are calibrated with spectroscopic observations of point sources, as we have explained in several prior papers (Lisse et al. 2006, 2007; Sitko et al. 2011). The IRS slits are narrow with respect to the PSF of the telescope, i.e., they do not encompass 100\% of the PSF at any wavelength and the fraction of the PSF encompassed varies with wavelength. Because the IRS instrument is calibrated with observations of spatially unresolved stars, the slit-losses may be ignored when working with calibrated observations of point sources. Ignoring the slit-loss does, however, affect the spectral shape of calibrated observations of sources larger than the PSF of the telescope (i.e., like what we have for SW1). We used the photometry from the IRAC 8.0 $\mu$m and MIPS 24.0 $\mu$m images, with an aperture of 9$''$.0 radius, to tie the spectral flux density values with the IRS spectrum, which can be seen in Figure 13.

\begin{figure}
\begin{centering}
\includegraphics[scale = 0.3]{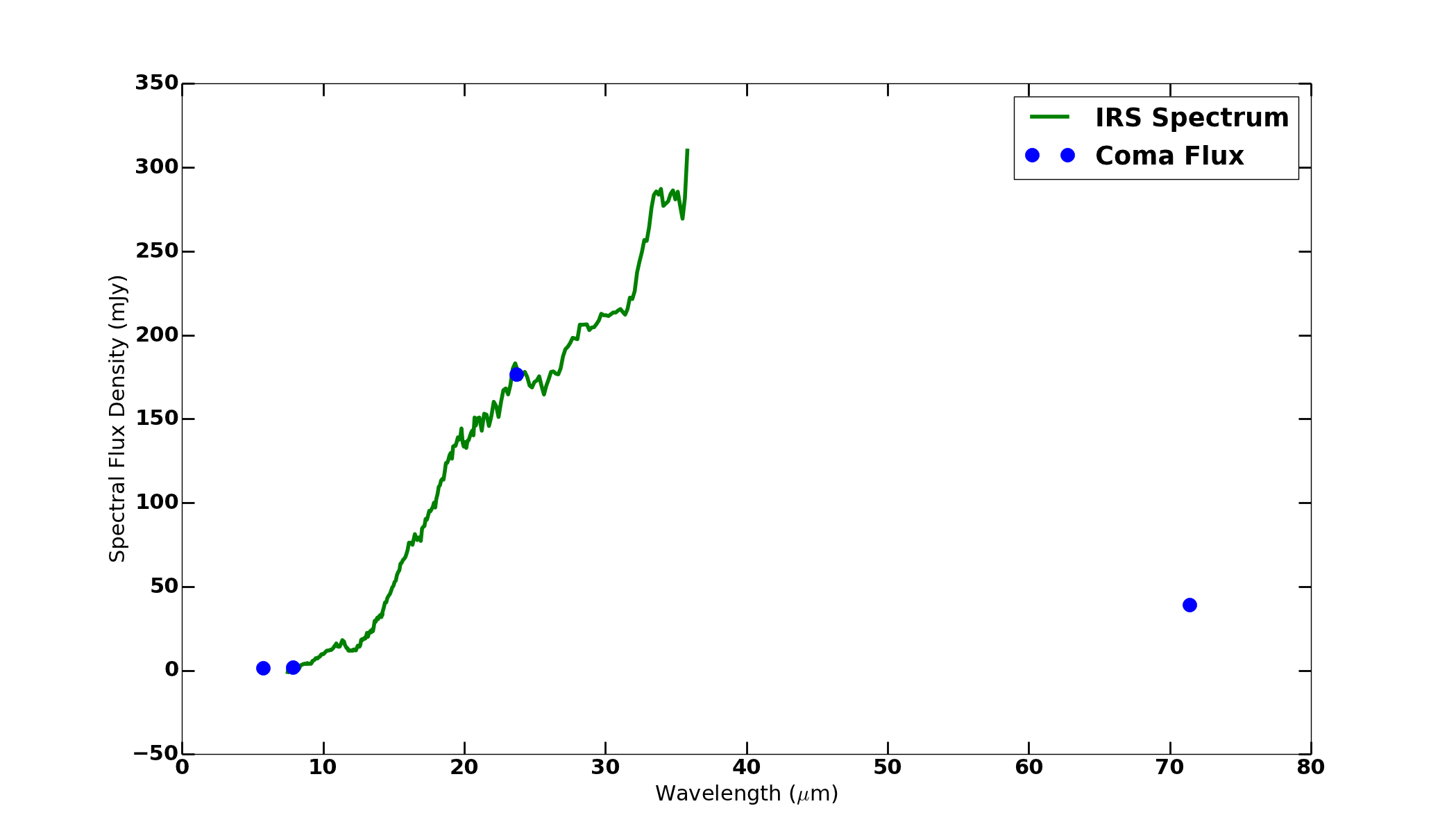}
\caption{Shown are the coma photometry measurements for the IRAC and MIPS images for a $9''.0$ radius aperture, along with the {\it Spitzer} IRS spectrum.}
\end{centering}
\end{figure}

\subsection{Spectral Modeling}

The infrared emission from a collection of dust near a comet nucleus has been explained in prior articles (Lisse et al. 1998, 2004, 2005, 2006, 2007; Reach et al. 2010; Sitko et al. 2011) and this procedure of spectral analysis is used here. The thermal emission from the dust coma is given by

$$F_{\lambda, \textrm{mod}} = \frac{1}{\Delta^2} \sum_i \int_0^{\infty} B_{\lambda}(T_i(a, r_h)) Q_{\textrm{abs}, i} (a, \lambda) \pi a^2 
    \frac{dn_i(r_h)}{da}da,$$

\noindent where $T$ is the temperature for a particle of radius $a$ and composition $i$ at heliocentric distance $r_h$, at a distance $\Delta$ from the 
observer, $B_{\lambda}$ is the blackbody radiance at wavelength $\lambda$, $Q_{\textrm{abs}}$  is the emissivity (emission efficiency) of the particle of composition $i$ at wavelength $\lambda$, $dn/da$ is the differential particle size distribution (PSD) of the emitted dust, and the sum is over all species of material and all sizes of particles for the dust. Spectral analysis consists of calculating the emission flux for a model collection of dust, and comparing the calculated flux to the observed flux. The emitted flux depends on the composition (location of spectral features), particle size (feature to continuum contrast), and the particle temperature (relative strength of short vs. long wavelength features). Prior to the modeling, the effects due to reflected sunlight and nucleus thermal emission need to be checked, and if necessary, removed. The IRS spectrum was divided by a Planck function, with temperature $T = 140$ K to remove the underlying blackbody profile of the dust grains. We excluded data below 8.0 $\mu$m due to the noise below this wavelength. The resulting relative emissivity spectrum along with a six component spectral model are shown in Figure 14.

\begin{figure}
\begin{centering}
\includegraphics[scale = 0.30]{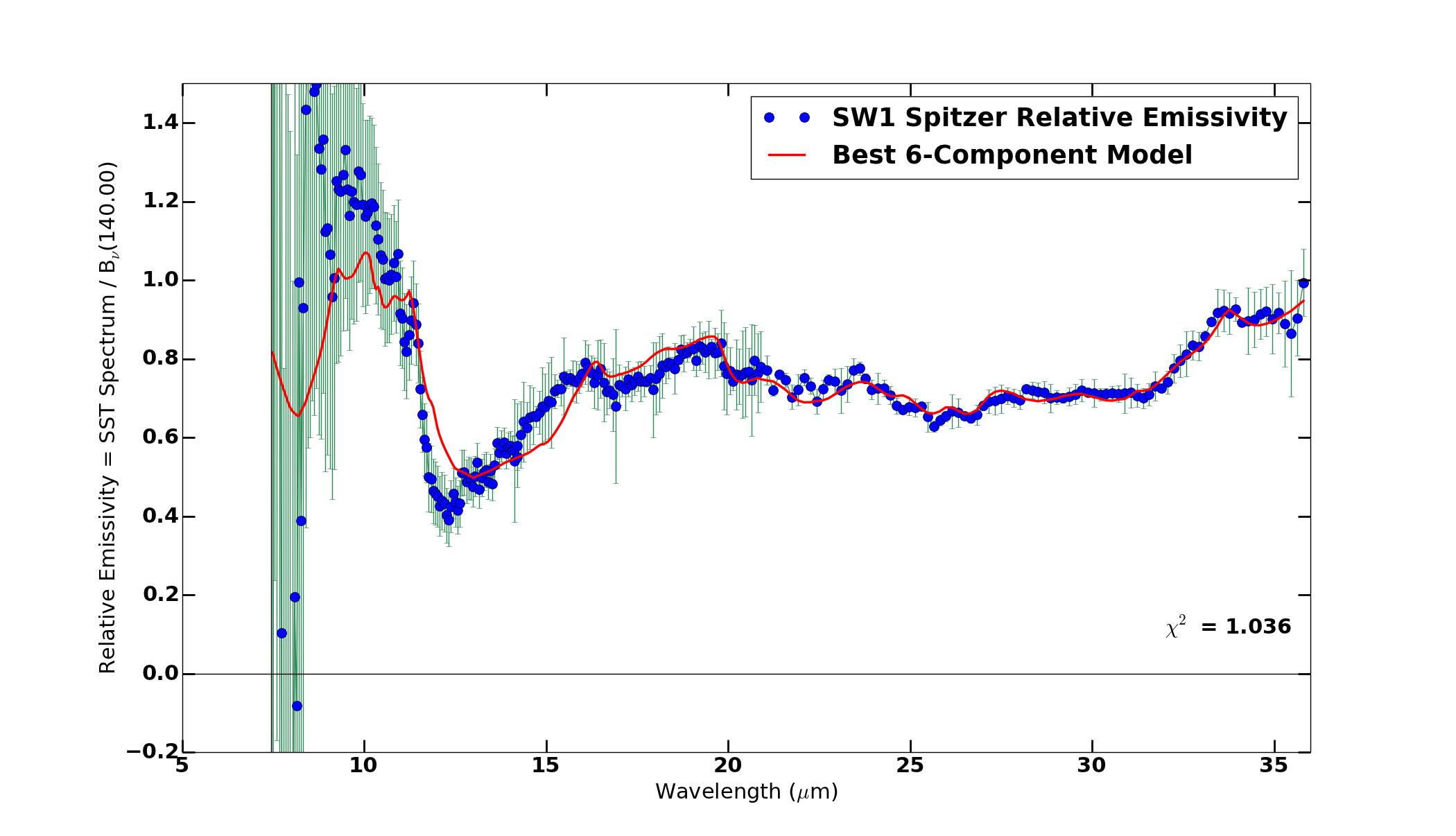}
\caption{IRS relative emissivity spectrum of SW1 and best-fit six component spectral model. Blue points show the extracted data from {\it Spitzer} IRS observations, along with green error bars for each data point. The reduced $\chi^2$ = 1.036 for model fit. The spectral decomposition of this model is shown in Figure 15. The model does a reasonable job of reproducing most of the broadband features, but there are wavelength regions (e.g. 12-16 $\mu$m) where the model deviates noticeably from the IRS observations. }
\end{centering}
\end{figure}

\begin{center}
\begin{table}
\caption{}
\resizebox{1.0\textwidth}{!}{
	\begin{tabular}{ l c c c c  c c }
		\multicolumn{7}{ c }{Composition of the Best-fit Model$^{\textrm{a}}$ to the {\it Spitzer} IRS SW1 Spectrum}\\
		\hline \hline
		Species 						&	Weighted		&	Density	  		&	M.W.	 	&	${N_{\textrm{moles}}}^{\textrm{b}}$	&	Model ${T_{\textrm{max}}}^{\textrm{c}}$ 	&	Model $\chi^2_{\nu}$ 	\\ 
									&	Surface Area	&	(g cm$^{-3}$)		&			&	(Relative)						&	(K)								&	(if not included)			\\ \hline
		Amorph Olivine (MgFeSiO$_4$)	&	0.515		&	3.6				&	172		&	1.08							&	140								&	62.10 				\\
		Forsterite (Mg$_2$SiO$_4$)		&	0.210		&	3.2				&	140		&	0.48							&	140								&	5.33 					\\		
		Diopside (CaMgSi$_2$O$_6$)		&	0.110		&	3.3				&	216		&	0.17							&	140								&	2.65 					\\
		Pyrrhotite (MgFeS)				&	0.140		&	4.5				&	 84		&	0.75							&	140								&	6.25 					\\
		Amoprh Carbon (C)				&	0.675		&	2.5				&	12		&	14.1							&	190								&	2.91 					\\
		Water ice (H$_2$O)				&	0.700		&	1.0				&	18		&	3.89							&	92								&	36.61				 \\
 \hline
	\end{tabular}

}
\caption*{
\resizebox{1.0\textwidth}{!}{
	\begin{tabular}{l}
		$^{\textrm{a}}$ Best-fit 6-component model with power-law particle size distribution $dn/da \sim a^{-3.75}$ (Note: using 40:60 Mg:Fe Amorph Olivine) . \\
		$^{\textrm{b}}$ ${N_{\textrm{moles}}}^{\textrm{b}} \sim$ density($i$)/molecular weight($i$) \textasteriskcentered normalized surface area($i$). Errors are $\pm$ 10\% for all species, except Amorph Carbon, which has $\pm$ 20\%.\\
		$^{\textrm{c}}$ Temperature of submicron- to micron-sized dust grains in K. Errors in temperature of dust are $\pm$ 5 K (2-$\sigma$).\\
	\end{tabular}
}	

}

\end{table}
\end{center}

The model and its component-by-component decomposition is shown in Table 3 and Figure 15. Note that while we limit ourselves to six components, it is possible to lower the best-fit $\chi^2$ by adding components. It may also be possible to lower the best-fitting $\chi^2$ by choosing a different set of six components. For these reasons, we emphasize that our result here for the dust composition is not necessarily unique. However we claim that the structure of the spectrum indicated that the six components we present in Figure 15 have at least a reasonable likelihood of actually being in the grains. Specifically: The amorphous olivine is clearly seen in the spectrum by the broad emission feature from 15-20 $\mu$m. There is some evidence also for it in the 9-12 $\mu$m emission feature although the spectral data are of poor signal-to-noise ratio there that we do not want to rely on that. Water ice is revealed in the spectrums the increase in emissivity past 25 $\mu$m. Diopside and forsterite are known by several emission features that seem to match such features in the spectrum itself long ward of 12 $\mu$m. Again, we do not want to rely on the features from those components at shorter wavelengths because the data are so uncertain. Pyrrhotite manifests itself as the broad emission around $\sim$24-33 $\mu$m and it has the effect of bumping up the entire spectrum at those wavelengths. Finally, amorphous carbon helps the model fitting short ward of 12 $\mu$m due to its upturn in flux relative to a 140 K blackbody there, but the high uncertainty of the data in this region makes it impossible to be definitive.

\begin{figure}
\begin{centering}
\includegraphics[scale = 0.30]{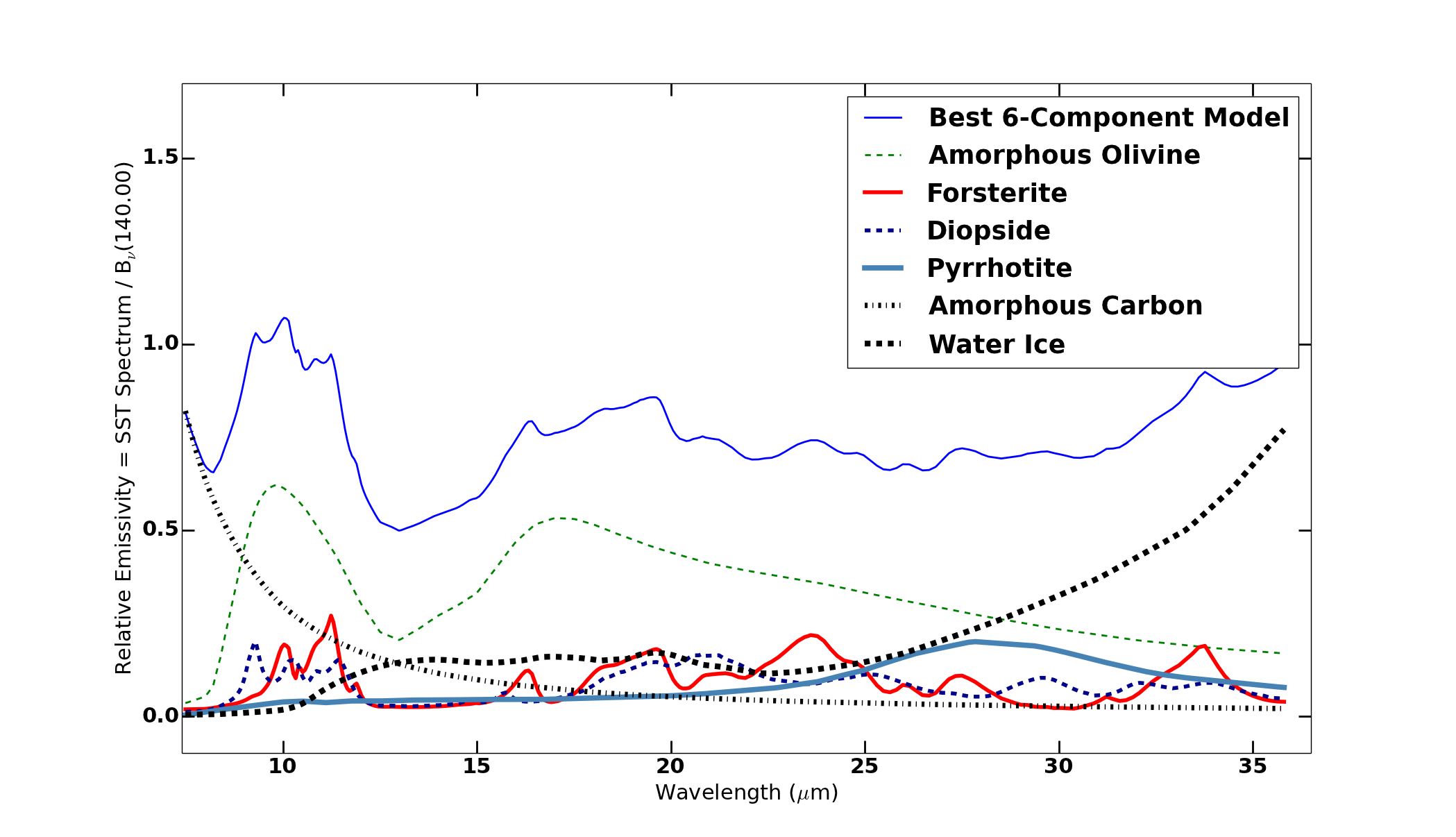}
\caption{Best-fit six component spectral model is shown with the individual six component lab spectra used for fitting. The model was built on a linear combination of the six individual lab spectra, varying the linear expansion coefficients until arriving at a minimum deviation from the IRS relative emissivity. When comparing the individual spectra with the IRS emissivity in Figure 14, the signatures for the compositional elements become noticeable (e.g. the broad absorption feature of amorphous olivine around 13.5 $\mu$m or the narrow features of forsterite from 17-34 $\mu$m).}
\end{centering}
\end{figure}

The derived dust temperature of 140 +/- 5 (2-$\sigma$) K is close to what is expected for submicron- to micron-sized Mg-Fe amorphous silicate and amorphous carbon grains at $\sim$ 5.2 to 5.6 AU from the Sun (e.g., Hanner et al. 1997). The derived particle size distribution (PSD) from the linear emissivity model is $dn/da \sim a^{-3.75 \pm 0.15}$ (2-$\sigma$), indicating a comet with a high relative abundance of small dust particles. The dust is amorphous-dominated, which has been seen in only one other (bright) comet, 73P (Sitko et al. 2011). Amorphous, primitive dust is unusual, and likely indicates that the material composing the refractory portion of SW1 was never heated to high temperatures, which is in contrast to crystalline silicate signatures observed in other comet dust for which we have good spectra (like 9P/Temple 1, Hale-Bopp (C/1995 O1), etc.). 

Mie models of the same {\it Spitzer} spectrum shows similar results, with a somewhat steeper PSD ($dn/da \sim a^{-3.9 \pm 0.20}$, but the difference between this and $dn/da \sim a^{-3.75 \pm 0.15}$ is within the normal range of modeling error). These results strongly suggest that the most likely Mie models are those with a high relative abundance of amorphous carbon, leading to a high ``superheat", large $T(\textrm{dust})/T(\textrm{blackbody})$ (Gehrz and Ney 1992), and low average dust albedo, as amorphous carbon grains are very efficient absorbers of optical solar energy (i.e., they are very dark), but only moderately good at re-radiating their energy in the thermal infrared.

\section{Conclusions and Discussion}

The reanalysis of archival {\it Spitzer} thermal observations of SW1 has led to new measurements of physical properties of the nucleus. The effective radius measurement of $R$ = 30.2 $^{-2.9}_{+3.7}$ km is within the 1-$\sigma$ uncertainties of several recent measurements (Stansberry et al. 2004; Stansberry et al. 2007), signifying a possible honing of the nuclear size measurement. The H$_{\textrm{v}}$ implied by the newly derived radius suggests a lower albedo for SW1. While the choice of geometric albedo and our determined value of the radius affects the H$_{\textrm{v}}$ value, we would like to emphasize that our choice of a geometric albedo p$_{\textrm{v}}$ = 0.04 does not have much of an effect on the radius derived from thermal photometry. The following radius measurements were arrived at when varying the geometric albedo: (p$_{\textrm{v}}$ = 0.06, R = 30.2 km), (p$_{\textrm{v}}$ = 0.08, R = 30.2 km), and (p$_{\textrm{v}}$ = 0.2, R = 30.1 km). All of the radius determinations are within the uncertainty of the value arrived at using p$_{\textrm{v}}$ = 0.04. If we choose to use p$_{\textrm{v}}$ = 0.04 then our measured radius leads to a value of H$_{\textrm{v}}$ = 10.24. Stansberry et al. (2004) used a value of H$_{\textrm{v}}$ = 11.0 (from Degewij et al. 1981; Meech et al. 1993) which with our radius determination yields a value of p$_{\textrm{v}}$ = 0.025, i.e., a lower albedo and a value that is consistent with some TNOs (Stansberry et al. 2008). The size of SW1 places it on the larger end of the Centaur size distribution. Considering the activity levels of SW1 at such a large heliocentric distance (e.g. Paganini et al. 2013) signifies a large supply of volatile material. If the orbit ever gets perturbed sending SW1 into the inner solar system it most likely will be an impressive sight.

An infrared beaming parameter of $\eta = 0.99$ $^{-0.19}_{+0.26}$ is near the middle of the $\eta$ distribution for other Centaurs. A collection of IR beaming parameter measurements for a database of 57 Jupiter family comets (JFC), 58 Centaurs, and 75 trans-Neptunian objects (TNO) (Bauer et al. 2013; Duffard et al. 2014; Fern\'andez et al. 2013; Fornasier et al. 2013; Lellouch et al. 2013; Mommert et al. 2012; P\'al et al. 2012; Santos-Sanz et al. 2012; Stansberry et al. 2008; Vilenius et al. 2012) is shown in Figure 16. The SW1 measurement is highlighted by a vertical bar on the histograms. The measured value of $\eta = 0.99$ for SW1 is towards the middle of each of the three small-body populations, as shown an in Figure 16. A value of $\eta = 0.99$ for SW1 could signify a surface with no radical topography (i.e. no high temperature regions causing infrared beaming). Alternatively, SW1 could have a low thermal inertia, resulting in little or no night side emission.

\begin{figure}
\begin{centering}
\includegraphics[scale = 0.3]{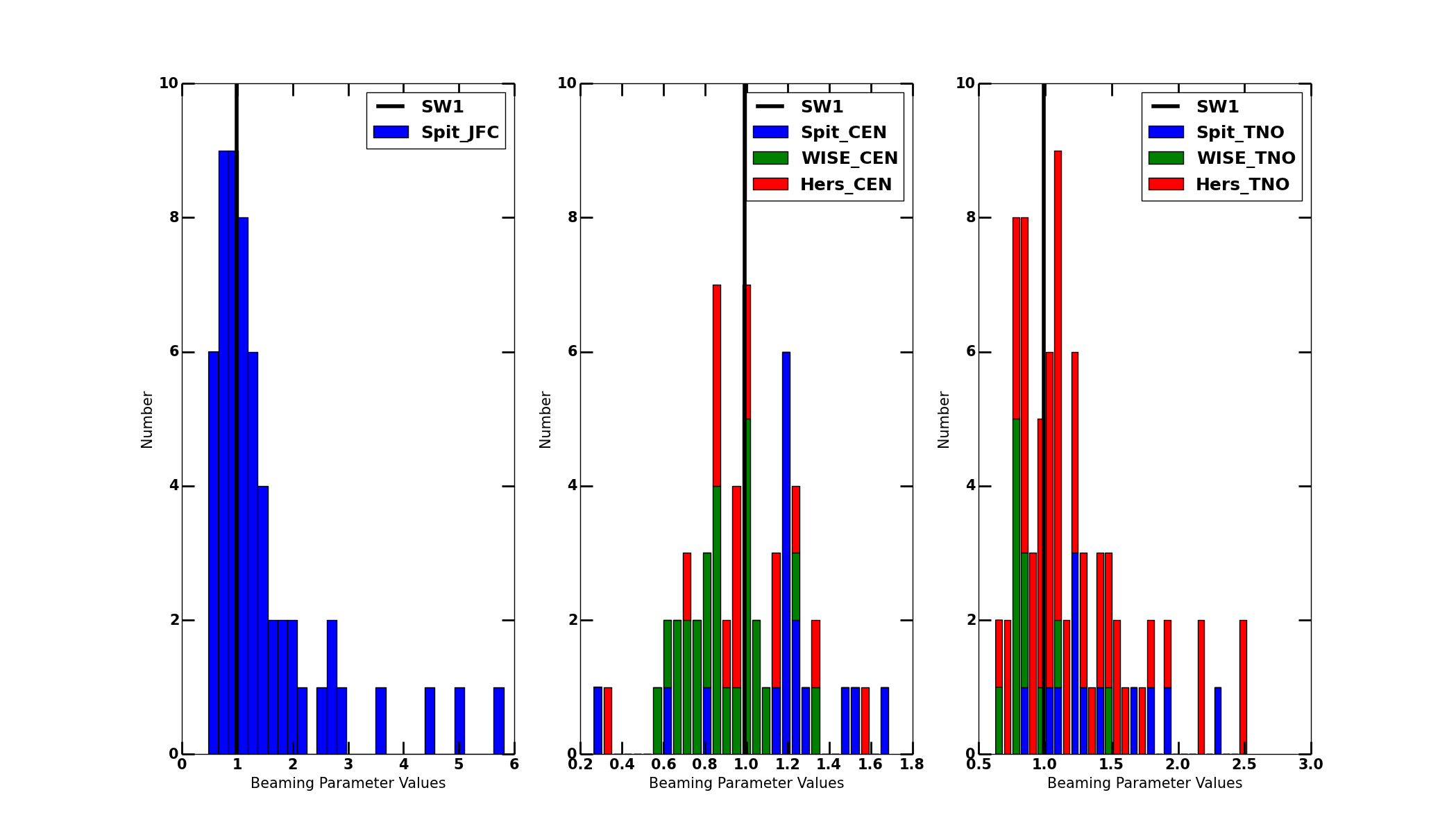}
\caption{Histograms of measured beaming parameters from current journal articles (see conclusion for list of beaming parameter sources). The vertical bar in each plot signifies our measured value of $\eta = 0.99$ for SW1. The left plot shows the distribution of 57 JFC and SW1's measured value is towards the mean of the distribution. All JFC $\eta$ values were acquired from {\it Spitzer} observations in SEPPCoN (Survey of the Ensemble Physical Properties of Cometary Nuclei) Fernandez et al. (2013). The middle plot shows SW1's placement in the ensemble of $\eta$ values for 58 Centaurs. Again SW1's measured value places it towards the middle of the distribution. Histogram colors represent the telescope used for observations (blue:{\it Spitzer} (Fern\'andez et al. 2013; Stansberry et al. 2008); green:{\it WISE} (Bauer et al. 2013); red:{\it Herschel} (Duffard et al. 2014; Fornasier et al. 2013; Lellouch et al. 2013; Mommert et al. 2012; P\'al et al. 2012; Santos-Sanz et al. 2012; Vilenius et al. 2012)). Plot to the right is similar, but with SW1's placement in the distribution of measured TNO $\eta$ values. Again SW1's measured value is towards the center of the distribution.}
\end{centering}
\end{figure}

While our size measurement is within uncertainties of several previous measurements, the infrared beaming parameter of each analysis is considerably different. Stansberry et al. (2004) used a value of $\eta = 0.62$. If we use their measured spectral flux density values, a fixed $\eta = 0.99$ value, and our model we arrive at a best fit radius value of 28.6 km. Unfortunately, a reduced $\chi^2 = 17.53$ signifies a poor modeling fit. This fit can be seen in Figure 17a. A similar analysis was performed on the Stansberry et al. (2007) spectral flux density measurements. A best-fit radius value of $R$ = 26.8 km was found with a smaller reduced $\chi^2 = 4.75$ value. This fit can be seen in Figure 17b. These values are shown to not undervalue previous measurements, but to emphasize the robustness in the measurements found in this analysis. Using archival {\it Spitzer} data products and newer more robust coma removal techniques, we have a size measurement that is in agreement with previous measurements, solidifying SW1's place as a large Centaur. More importantly, our modeling has resulted in a beaming parameter value close to the average of a sampling of JFCs, Centaurs and TNOs.

\begin{figure}[h!]

	\centering
	\begin{subfigure}[b]{0.9\textwidth}
	
		\includegraphics[scale = 0.3]{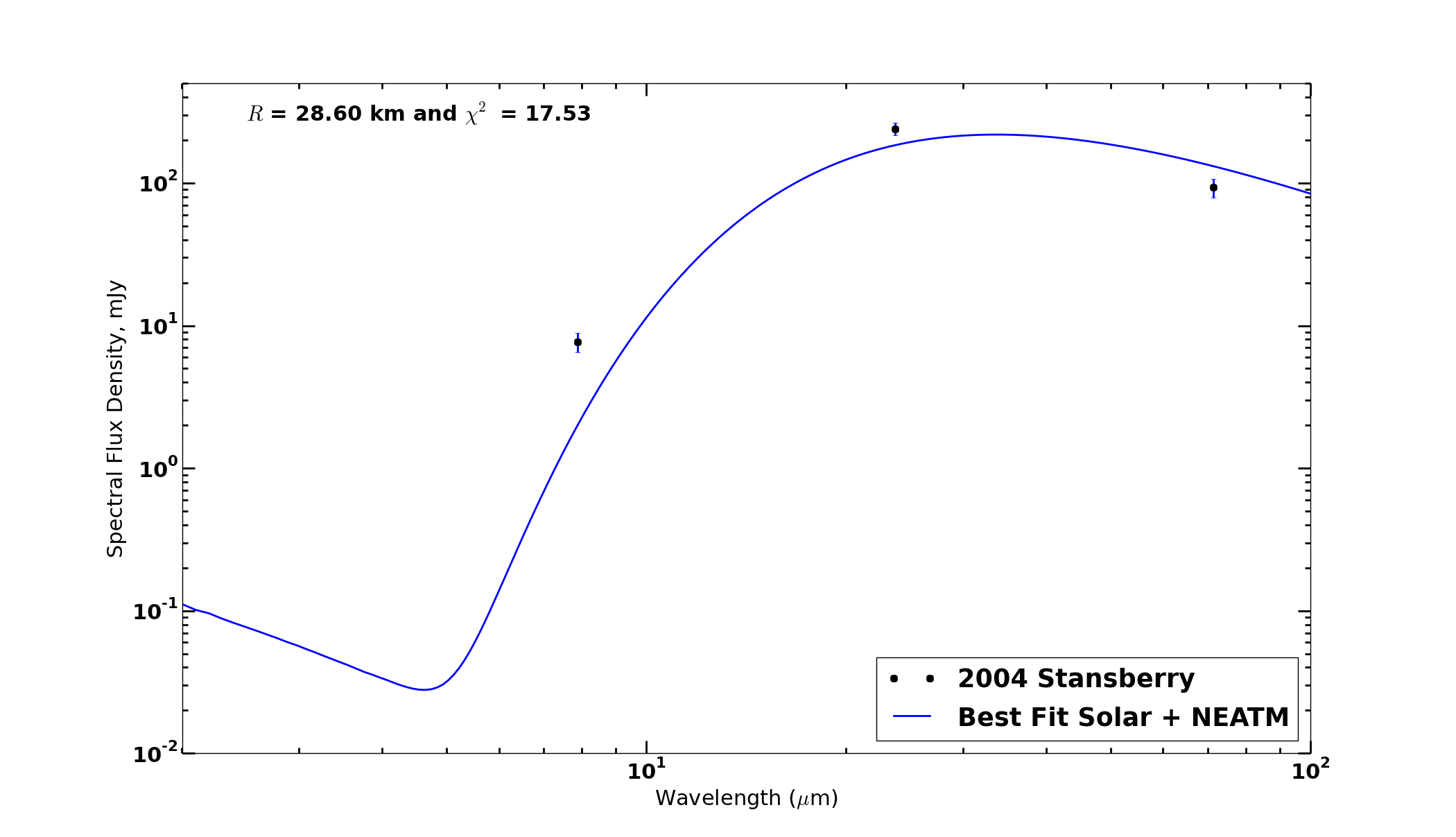}
		\caption{}
	
	\end{subfigure}

	\begin{subfigure}[b]{0.9\textwidth}
	
		\includegraphics[scale = 0.3]{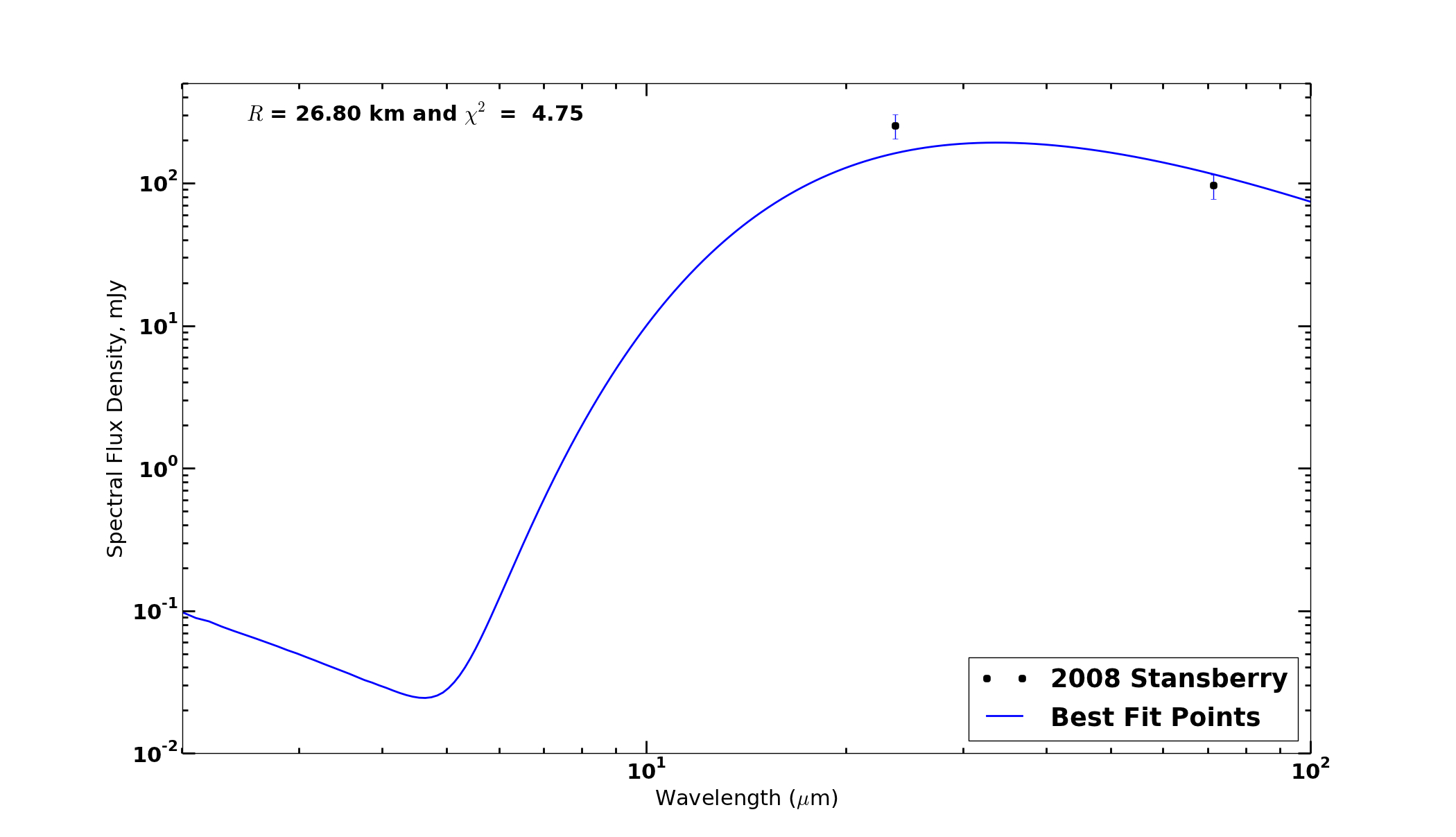}
		\caption{}
	
	\end{subfigure}
\caption{Modeling fits to the Stansberry et al. (2004) and (2008) spectral flux density measurements. }

\end{figure}

We were unable to find constraints for the 5.8 $\mu$m geometric albedo of SW1 with the thermal observations alone (no coinciding observations of SW1 in the visible were available). Thermal modeling resulted in a value of 0.5 $\pm$ 0.5 which shows a high degree of uncertainty.  While we do not claim an infrared geometric albedo measurement of 0.5, we would like to point out albedo measurements in this wavelength region are lacking from literature. Infrared albedos may be higher than the usual 0.04 values found for a geometric albedo in the visible. Although the uncertainty in our measurement is high as well, we would like to show that this magnitude of infrared albedo is not physically out of the question. A normalized reflectivity gradient $S' = 14.94 \pm 1.09$ [\% (1000 \r{A})$^{-1}$] (Duffard et al. 2014) and a V-band albedo of 0.04 would yield an albedo of 0.31 $\pm$ 0.10 in the near-infrared. Note that the normalized reflectivity gradient is a simplification and assumes a linear relation between albedo change and wavelength, which is truly not the case for mid-infrared albedos. It is mentioned here to just show the reader that the albedos of Centaurs tend to increase towards the infrared and that a value of 0.04 is not valid at these longer wavelengths. The high uncertainty in the derived IR albedo also means that SW1 may have a much lower IR albedo that 0.5.

Using a linear combination of silicate, amorphous carbon, water ice, and metal sulfide laboratory emissivities, shown to successfully fit and explain the 9P/Tempel 1, Hale-Bopp (C/1995 O1), 17P/Holmes, and 73P/Schwassmann-Wachmann 3 comet dust spectra (Lisse et al. 2007; Reach et al. 2010; Sitko et al. 2011), we find the fit for the {\it Spitzer} SW1 spectrum shown in Figure 14. Containing 281 spectral measurements, this spectrum is much more highly constraining on the possible dust composition and size models. Only fits containing appreciable amounts of silicates, water ice, and amorphous carbon fit the data well.

\underline{To summarize:}
\begin{itemize}

  \item The measured value of $R$ = 30.2 $^{+3.7}_{-2.9}$ km places SW1 on the larger end of the Centaur size distribution. This value has been shown to be consistent (within 1-$\sigma$) of several previous measurements.

  \item An infrared beaming parameter value, $\eta=0.99$, measured for SW1 is towards the middle of distributions of measured $\eta$ values for JFCs, Centaurs, and TNOs.  

  \item The 5.8 $\mu$m infrared albedo was measured to be 0.5 $\pm$ 0.5, but with such a large uncertainty in this value it is still unclear how the reflectivity truly behaves in this wavelength regime. It was shown, however, that using a normalized reflectivity gradient for SW1 and a standard visible albedo for comets that a high ($>$ 0.30) albedo can be achieved.  
  
  \item Finally, using {\it Spitzer} IRS observations a dust composition model was found to show SW1's dust is dominated by amorphous minerals, with possible signatures for more thermally processed minerals. 

\end{itemize}

\section*{Acknowledgements}

This work is based on observations made with the {\it Spitzer} Space Telescope, which is operated by the Jet Propulsion Laboratory, California Institute of Technology under a contract with NASA. This research made use of Tiny Tim/{\it Spitzer}, developed by John Krist for the {\it Spitzer} Science Center. The Center is managed by the California Institute of Technology under a contract with NASA. We acknowledge funding support from NASA's Outer Planets Research program (grant NNX12AK50G). 

We would also like to thank Dr. Beatrice Mueller, Dr. Michael Kelley and our two anonymous reviewers for their comments and suggestions during the preparation of this article.


\end{document}